\documentclass[reprint, amsmath, amssymb,aps,prb,superscriptaddress]{revtex4-2}
\usepackage{graphicx} 
\usepackage{bm}
\usepackage{braket}
\usepackage{mathtools}
\usepackage{color}
\usepackage{here}
\usepackage{comment}
\usepackage{physics}
\usepackage{amsthm}
\usepackage{enumitem}

\newcommand{\so}[1]{{\color{orange}{#1}}}
\newcommand{\soo}[1]{{\color{red}{#1}}}

\newcommand{\soooo}[1]{{\color{orange}{#1}}}
\newcommand{\imi}{{\rm i}}

\renewcommand{\so}{}
\renewcommand{\soo}{}
\renewcommand{\soooo}{}

\usepackage[dvipsnames]{xcolor}
\definecolor{DarkGreen}{rgb}{0.0, 0.5, 0.0}
\usepackage[colorlinks,citecolor=DarkGreen,linkcolor=blue,linktocpage]{hyperref}

\def \epsilonnovar {\epsilon}

\def \fbar {\bar{f}}

\def \fplus {f_k^\dagger}
\def \fminus {f_k}
\def \rhofunc {F}

\begin{document}
\title{Exact spectrum and anomalous relaxation in the open disorder-free Sachdev-Ye-Kitaev system}
\author{Soshun Ozaki}
 \affiliation{Department of Basic Science, The University of Tokyo, Komaba, Meguro-ku, Tokyo 153-0041, Japan}
 \affiliation{Max Planck Institute for Solid State Research, Heisenbergstrasse 1, D-70569 Stuttgart, Germany}
 \affiliation{Department of Physics, Chuo University, Kasuga, Bunkyo-ku, Tokyo 112-8551, Japan}
\author{Hironobu Yoshida}%
\affiliation{Department of Physics, 
The University of Tokyo, 
Hongo, Bunkyo-ku, Tokyo 113-0033, Japan}
\affiliation{Nonequilibrium Quantum Statistical Mechanics RIKEN Hakubi Research Team,
RIKEN Pioneering Research Institute (PRI), Wako, Saitama 351-0198, Japan}

\affiliation{Department of Physics, Princeton University, Princeton, New Jersey 08544, USA}

\date{\today}
\author{Hosho Katsura}%
\affiliation{Department of Physics, 
The University of Tokyo, 
Hongo, Bunkyo-ku, Tokyo 113-0033, Japan}
\affiliation{Institute for Physics of Intelligence, The University of Tokyo, 
Hongo, Bunkyo-ku, Tokyo 113-0033, Japan}
\affiliation{Trans-Scale Quantum Science Institute, The University of Tokyo, 
Hongo, Bunkyo-ku, Tokyo 113-0033, Japan}
\date{\today}

\begin{abstract}
We study a disorder-free variant of the Sachdev-Ye-Kitaev (SYK) model with dissipation within the Gorini-Kossakowski-Sudarshan-Lindblad formalism. By utilizing the integrability of the clean SYK model, we derive an exact solution 
in a spectrum-resolved form, i.e., the eigenvalues and corresponding projection superoperators of the Liouvillian for arbitrary system size $N$. We determine the scaling of the gap that governs the long-time decay of the two-point correlation functions. Importantly, the gap does not vanish in the dissipationless limit when the thermodynamic 
limit is taken first, despite the integrability of the model. This phenomenon, known as anomalous relaxation, suggests a possible connection with chaotic dynamics and quantum Ruelle-Pollicott resonances.
We also find several spectral features, such as transitions in the Liouvillian spectrum from complex to real 
eigenvalues with increasing dissipation strength, as well as the convergence of the dissipative form factor to the spectral form factor in the dissipationless limit. These findings indicate that the present model offers a useful platform for exploring nontrivial open dynamics of many-body quantum systems.
\end{abstract}

\maketitle
\section{Introduction}
Understanding the origin and universality of quantum chaos in interacting many-body systems remains a central problem in modern theoretical physics.
The Sachdev-Ye-Kitaev (SYK) model has played a pivotal role in this context, offering an exceptional platform to study strongly interacting systems that exhibit maximal chaos in the large-$N$ limit and universal low-energy dynamics \cite{sy1993,kitaev,trunin,rosenhaus2019introduction,chowdhury2022sachdev}.
Its analytical tractability has enabled systematic studies of dynamical chaos diagnostics, including spectral correlations and out-of-time-order correlators (OTOCs) \cite{Cotler2017,Maldacena2016prd,Kobrin2021}.

A natural question arising from the properties of the SYK model is to what extent quantum chaos is tied to randomness and nonintegrability.
To address this issue, several deterministic variants of the SYK model have been introduced \cite{gurau2017complete,klebanov2017uncolored,krishnan2018contrasting,iyoda2018pra,iyoda2018,witten2019syk,Wang2020,Balasubramanian2021,Claps,gorsky2025theta,biggs2026}.
Among them, the SYK model with uniform \so{four-body} couplings \cite{Lau2021,ozaki-katsura,fukai-katsura2026},
\begin{equation}
    H_4=-\sum_{1\leq i<j<k<l \leq N}\gamma_i \gamma_j \gamma_k \gamma_l,
    \label{eq:H4}
\end{equation}
serves as a simple yet particularly intriguing example, where $\gamma_i$'s ($i=1,\dots, N$) are Hermitian Majorana operators satisfying $\{\gamma_i,\gamma_j\}=2\delta_{ij}$.
This so-called clean SYK model is integrable for arbitrary $N$ and therefore non-chaotic in the conventional sense of closed quantum systems.
However, a previous study has revealed that it exhibits anomalous operator dynamics and OTOC behavior reminiscent of chaotic systems, with the caveat that the duration is short and independent of $N$ \cite{ozaki-katsura}.
This observation highlights nontrivial operator dynamics even in integrable quantum many-body systems.

In parallel, recent progress in open quantum systems has drawn attention to the Liouvillian spectrum of the Gorini-Kossakowski-Sudarshan-Lindblad (GKSL) equation \cite{gksl1,gksl2} as a powerful probe of irreversibility of purely unitary dynamics in closed quantum many-body systems~
\cite{sa_lindbladian_2022,garcia_keldysh_2023,garcia-garcia_UniversalityItsLimits_2023,mori_liouvillian_2024,shackleton_ExactlySolvableDissipative_2024,yoshimura_RobustnessQuantumChaos_2024a,jacoby_SpectralGapsLocal_2025a,sa_ExactlySolvableDissipative_2025,yoshimura_TheoryIrreversibilityQuantum_2025a}.
The relaxation rate of chaotic open quantum many-body systems remains finite when the thermodynamic limit is taken before the weak-dissipation limit.
This anomalous relaxation has been discussed in connection with chaotic dynamics and quantum Ruelle–Pollicott resonances~\cite{manderfeld_ClassicalQuantumTime_2001,prosen_RuelleResonancesQuantum_2002,garcia-mata_ClassicalDecaysDecoherent_2003,nonnenmacher_SpectralPropertiesNoisy_2003,prosen_RuelleResonancesKicked_2004,prosen_ChaosComplexityQuantum_2007,garcia_keldysh_2023,mori_liouvillian_2024,znidaric_MomentumdependentQuantumRuellePollicott_2024b,teretenkov_PseudomodeExpansionManybody_2025, Yamashika2026}.
Analytical results confirming this anomalous relaxation have been obtained for $q$-body SYK models in the large $N$ limit~\cite{garcia_keldysh_2023,kulkarni2022prb}, as well as for random quantum circuits~\cite{yoshimura_RobustnessQuantumChaos_2024a,jacoby_SpectralGapsLocal_2025a,yoshimura_TheoryIrreversibilityQuantum_2025a}
with dissipation.
Various aspects of dissipative dynamics in SYK-related models, including Liouvillian spectra, dissipative form factor (DFF), and \soooo{operator growth} have also been investigated~\cite{kulkarni2022prb,kawabata2023prb,kawabata2023prxq, garcia-garcia_UniversalityItsLimits_2023,almeida2026,Zheng2026prd,Bhattacharjee2023jhep, Bhattacharjee2024jhep}.

However, when the Hamiltonian has no randomness, the emergence of anomalous relaxation has been mainly explored numerically~\cite{mori_liouvillian_2024}, and analytical results remain limited~\cite{shackleton_ExactlySolvableDissipative_2024,sa_ExactlySolvableDissipative_2025}. 
Investigating analytically tractable models exhibiting anomalous relaxation is therefore valuable for clarifying its origin and mechanism.

From a theoretical standpoint, analyzing interacting GKSL dynamics is notoriously difficult, since the Liouvillian superoperator acts on a $D^2$-dimensional operator space for a Hilbert space of dimension $D$.
Consequently, exact analytical and controlled numerical results for interacting GKSL dynamics remain scarce and are typically restricted to highly constrained settings~\cite{znidaric1,znidaric2, Medvedyeva,shibata-katsura,kulkarni2022prb,kawabata2023prb,kawabata2023prxq,Prosen2009,Prosen_2008_njp,Nakagawa_2021_prl,Pozsgay}.
Therefore, identifying analytically tractable models of interacting dissipative systems remains a crucial challenge in this field. One promising direction is to study integrable models
in the presence of dissipation. Such models provide a useful testing ground for understanding anomalous relaxation.

In this paper, we study the clean SYK model with dissipation as a platform for this purpose, since it exhibits 
nontrivial quantum dynamics despite its integrability. We derive the exact Liouvillian spectrum for the clean SYK model in an open quantum setting and demonstrate that this model exhibits anomalous relaxation.
Exploiting the exact solvability and symmetry structure of the model, we analytically and numerically determine the scaling of the gap that governs the long-time decay of two-point correlation functions in the thermodynamic limit.
Thus, the present model offers a controlled setting for investigating anomalous relaxation in interacting open quantum systems.
In addition, we identify transitions of the Liouvillian spectrum from complex to real eigenvalues with increasing dissipation strength, associated with exceptional-point structures 
ubiquitous in non-Hermitian systems.

This paper is organized as follows.
In Sec.~\ref{sec:cleanSYK}, we review the exact solution of the clean SYK model.
In Sec.~\ref{sec:spectraldecomposition}, we derive the spectral decomposition of the Liouvillian. The DFF is also computed.
In Sec.~\ref{sec:examples}, we illustrate the spectral decomposition with several examples.
In Sec.~\ref{sec:correlation}, we analyze the dissipative dynamics of two-point correlation functions.
Finally, we summarize our results in Sec.~\ref{sec:summary}.
Details of the calculations and supporting data are provided in Appendices A--E.

\section{The clean SYK model\label{sec:cleanSYK}}
In this section, we briefly summarize the exact solution 
of $H_4$ (Eq.~\eqref{eq:H4}) \cite{Lau2021,ozaki-katsura}.
To solve $H_4$, we introduce the \textit{two-body} clean SYK model,
\begin{align}
    H_2=\imi \sum_{1\leq i<j \leq N}\gamma_i \gamma_j,
\end{align}
which is related to $H_4$ via 
\begin{equation}
 H_4=\frac{1}{2}(H_2)^2 -\frac{1}{4}N(N-1).
     \label{eq:h2h4relation}
\end{equation}
By Fourier transforming the Majorana fermions, we can define $\frac{N}{2}$ complex fermions,
\begin{align}
    &\fplus=\frac{1}{\sqrt{2N}} \sum_{j=1}^N e^{- \imi (j-1)\theta_k}\gamma_j,  \\
    &\fminus=\frac{1}{\sqrt{2N}} \sum_{j=1}^N e^{ \imi (j-1)\theta_k}\gamma_j,
\end{align}
with $\theta_k=(2k-1)\pi/N$ and $k=1,\dots,N/2$. These fermions satisfy
\begin{align}
    &\{\fminus,f_{k'}^\dagger\}=\delta_{kk'}, \nonumber \\
    &\{\fminus,f_{k'}\}=\{\fplus,f_{k'}^\dagger\}=0.
\end{align}
Using these complex fermions, $H_2$ is diagonalized as
\begin{equation}
    H_2=\sum_{k=1}^{N/2} \epsilonnovar_k \left( \fplus \fminus -\frac{1}{2} \right), \quad \epsilonnovar_k=2 \cot\frac{\theta_k}{2}.
    \label{eq:h2}
\end{equation}
Combining Eqs.~\eqref{eq:h2} and \eqref{eq:h2h4relation}, the energy eigenvalues of $H_4$ are obtained as
\begin{equation}
    E(n_1, \dots, n_{N/2})=\frac{1}{8}\bigg(\sum_{k=1}^{N/2}\epsilon_k (-1)^{n_k+1}\biggr)^2-\frac{N(N-1)}{4},
\end{equation}
where $n_k=0,1$ denotes the occupation number of the $k$th fermion. This result indicates that the quasiparticle picture valid for $H_2$ breaks down for $H_4$.

\section{Master equation and its Spectral decomposition \label{sec:spectraldecomposition}}
Next, we consider the following GKSL-type master equation for 
the clean SYK system with dissipation:
\begin{subequations}
\begin{align}
    \dot{\rho}&=\mathcal{L}[\rho], 
    \\
    \mathcal{L}[X]&=-\imi [H_4,X]+ \sum_{j=1}^N \left(L_j X L_j^\dagger -\frac{1}{2}\{L_j^\dagger L_j,X\}\right), \label{eq:lindbladian}\\
    L_j&=\sqrt{\Gamma}\gamma_j,
\end{align}
\label{eq:master}
\end{subequations}
where $\Gamma$ $(\geq0)$ is the strength of the dissipation.
The goal of this section is to obtain the spectral decomposition of the Liouvillian $\mathcal{L}$, which is defined by
\begin{align}
    &\mathcal{L}=\sum_m \lambda_m \Pi_m, \quad \mathcal{L} \Pi_m=\Pi_m \mathcal{L}=\lambda_m \Pi_m , \nonumber \\
    &\Pi_m \Pi_n=\delta_{mn}\Pi_m, \quad \sum_m\Pi_m=\mathcal{I},
    \label{eq:spectralres}
\end{align}
where $\lambda_m$ is the $m$th eigenvalue, $\Pi_m$ is the corresponding projection superoperator, and $\mathcal{I}$ is the identity superoperator. 

First, we briefly discuss the readily identifiable eigenvalue and its corresponding 
projection superoperators. Since all the jump operators are Hermitian, $\Pi_0[\,\bullet\,] =2^{-N/2} (\operatorname{Tr} [\,\bullet\, ]) I$ is an eigenprojection with eigenvalue $\lambda_0=0$, and therefore it yields a steady state. Moreover, this steady state is unique: every initial state relaxes to the infinite-temperature state 
$\rho_\infty=2^{-N/2}I$. This follows from Spohn's theorem \cite{spohn_approach_1976,spohn_algebraic_1977,frigerio_quantum_1977,frigerio_stationary_1978,spohn_kinetic_1980,nigro_uniqueness_2019,yoshida_2024,zhang_CriteriaDaviesIrreducibility_2024a,fagnola_IrreducibilityQuantumMarkov_2025a,hamazaki_IntroductionMonitoredQuantum_2025a}, which states that if the jump operators $L_j$ generate the full operator algebra on the Hilbert space under multiplication, addition, and scalar multiplication, then the steady state is unique. In our setting, the jump operators are the set of Majorana operators $L_j=\sqrt{\Gamma}\gamma_j$, which indeed generate the full operator algebra. Note that this argument involves only the Lindblad operators and not the Hamiltonian. Consequently, with our choice of Lindblad operators, the steady state is unique for any Hamiltonian, including the original SYK model.

Next, we determine the general eigenvalues and corresponding projection superoperators.
After the Fourier transform, we can rewrite the jump terms in $\mathcal{L}$ (Eq.~\eqref{eq:lindbladian}) 
in terms of the complex fermions,
\begin{align}
    \sum_{i=1}^N \gamma_i \rho \gamma_i 
    =2\sum_{k=1}^{N/2} \big[ \fplus \rho \fminus + \fminus \rho \fplus].
\end{align}
To eliminate the commutator and anticommutator in $\mathcal{L}$ (Eq.~\eqref{eq:lindbladian}), we perform the transformation \cite{honda-nakazato, ozaki-nakazato}
\begin{align}
    \tilde{\rho}(t)=e^{N\Gamma t} e^{\imi H_4 t}\rho(t) e^{-\imi H_4 t}.
\end{align}
As a result, Eq.~\eqref{eq:master} becomes
\begin{align}
    \dot{\tilde{\rho}}(t)=2\Gamma \sum_{k=1}^{N/2} \mathcal{K}_k(t) \big[\tilde{\rho}(t) \big],
\end{align}
where $\mathcal{K}_k(t)$ is a time-dependent superoperator whose action on a density matrix $\rho$ is
\begin{align}
    \mathcal{K}_k(t)[\rho] &=  e^{\imi \epsilon_k H_2 t} \fplus \rho \fminus e^{-\imi \epsilon_k H_2 t}+ e^{-\imi \epsilon_k H_2 t} \fminus \rho \fplus e^{\imi \epsilon_k H_2 t} \nonumber \\
    &=e^{\imi \sigma_k \epsilon_k H_2 t} (\fplus \rho \fminus + \fminus \rho \fplus) e^{-\imi \sigma_k \epsilon_k H_2 t},
\end{align}
with $\sigma_k = 2f_k^\dagger f_k-1$.
In deriving this expression, we have used the time evolution of complex fermions in the Heisenberg representation \cite{ozaki-katsura}: 
\begin{align}
    &e^{\imi H_4 t}\fminus e^{-\imi H_4 t} =e^{-\frac{\imi}{2} \epsilon_k^2 t } e^{-\imi \epsilon_k H_2 t}\fminus=\fminus e^{-\imi \epsilon_k H_2 t} e^{\frac{\imi}{2} \epsilon_k^2 t}, \nonumber \\
    &e^{\imi H_4 t}\fplus e^{-\imi H_4 t} =e^{-\frac{\imi}{2} \epsilon_k^2 t } e^{\imi \epsilon_k H_2 t}\fplus=\fplus e^{\imi \epsilon_k H_2 t} e^{\frac{\imi}{2} \epsilon_k^2 t}. \label{eq:f_time_evo}
\end{align}
Using Eq.~\eqref{eq:f_time_evo}, we obtain
\begin{align}
    [\mathcal{K}_{k_1}(t),\mathcal{K}_{k_2}(t')]=0 \quad \text{for} \,k_1\neq k_2.
    \label{eq:kcomm}
\end{align}
This commutation relation gives the formal solution of Eq.~\eqref{eq:master}:
\begin{align}
    \rho(t)=e^{-N\Gamma t}e^{-\imi H_4 t} \biggl(\prod_{k=1}^{N/2} \Lambda_k(t)\biggr)[\rho(0)]e^{\imi H_4 t}\label{eq:formalsol}
\end{align}
with
\begin{align}
    \Lambda_k(t) = T\exp\biggl[2\Gamma\int \mathcal{K}_k(t) dt\biggr],
    \label{eq:timeordered}
\end{align}
where $T$ denotes the time-ordering operator.
We note that $[\Lambda_k(t),\Lambda_{k'}(t)]=0$ for $k \neq k'$, which follows from Eq.~\eqref{eq:kcomm}.

To obtain an explicit expression for Eq.~\eqref{eq:formalsol}, we represent the density matrices 
in terms of the creation and annihilation operators of fermions.
We consider the complete orthonormal system (CONS) for the single mode operator space at momentum $k$,
\begin{align}
    \langle a_k b_k \rangle = 
    \begin{dcases}
        \bar{f}_k & (a_k,b_k)=(0,0) \\
        \bar{f}_k^\dagger &  (a_k,b_k)=(0,1)\\
        \bar{f}_k \bar{f}_k^\dagger & (a_k, b_k)=(1,0)\\
        \bar{f}_k^\dagger \bar{f}_k & (a_k, b_k)=(1,1)
    \end{dcases},
\end{align}
where $\fbar_k=f_k(-1)^Q$ and $\fbar_k^\dagger= (-1)^Q f_k^\dagger$.
Here, $Q=\sum_{k=1}^{N/2}f^\dagger_kf_k$ is the fermion number operator.
These modified fermion operators satisfy
$[\bar{f}_k^\dagger, f_{k'}^\dagger]=[\bar{f}_k, f_{k'}]=[\bar{f}_k,f_{k'}^\dagger]=[\bar{f}_k^\dagger, f_{k'}]=0$  for $k\neq k'$ \cite{vernier}.

Using these fermions, we construct the many-body CONS with $N$ binary variables $\bm a=(a_1,\dots,a_{N/2})$ and  ${\bm b}=(b_1, \dots, b_{N/2})$, with $a_k, b_k \in \{0,1\}$, as
\begin{equation}
    \prod_{k=1}^{N/2} \langle a_k b_k\rangle.
\end{equation}
\so{This many-body CONS is suitable for writing down the formal solution~Eq.~\eqref{eq:formalsol} in a closed form. Applying $\Lambda_k(t)$ to a single basis element, we obtain}
\begin{widetext}
\begin{align}
    \Lambda_k(t) \prod_{k'=1}^{N/2}\langle a_{k'} b_{k'}\rangle =&\sum_{c_k,d_k}
    [V_k^{a_k}({\bm a},{\bm b})]_{c_k d_k}  [V_k^{-a_k}({\bm a},{\bm b})]_{d_k b_k} \exp\biggl[a_k (-1)^{d_k}\eta_k({\bm a},{\bm b}) t\biggr]  
    \nonumber \\
    &\times\exp(\frac{\imi}{2}a_k \sigma_k \epsilon_k H_2 t) \biggl[\prod_{k'=1}^{k-1} \langle a_{k'} b_{k'}\rangle\biggr]
    \langle a_k c_k \rangle
    \biggl[\prod_{k'= k+1}^{N/2} 
    \langle a_{k'} b_{k'}\rangle\biggr]  \exp(-\frac{\imi}{2} a_k \sigma_k \epsilon_k H_2 t),
    \label{eq:single_k_timeevo}
\end{align}
where
\begin{align}
    V_k({\bm a},{\bm b})= \so{
    \begin{pmatrix}
        [V_k(\bm a,\bm b)]_{00} & [V_k(\bm a,\bm b)]_{01} \\
        [V_k(\bm a,\bm b)]_{10} & [V_k(\bm a,\bm b)]_{11} 
    \end{pmatrix}
    =}
    \begin{pmatrix}
        \frac{\imi}{2} \alpha_k({\bm a},{\bm b}) + \eta_k({\bm a},{\bm b}) & -2\Gamma \\
        2\Gamma & \frac{\imi}{2} \alpha_k({\bm a},{\bm b})+\eta_k({\bm a},{\bm b}) 
    \end{pmatrix},
    \label{eq:vmatrix}
\end{align}
and
\begin{align}
    &\alpha_k({\bm a},{\bm b})= -\epsilon_k \sum_{k'\neq k} \epsilon_{k'}(1-a_{k'})(-1)^{b_{k'}},  \label{eq:alpha} \\
    &\eta_k({\bm a},{\bm b}) 
    = \sqrt{4 \Gamma^2-\frac{1}{4}\alpha_k({\bm a},{\bm b})^2}
    \label{eq:evplusmain}.
\end{align}
We note that $V^{-1}_k({\bm a},{\bm b})$ and $V_k^0({\bm a},{\bm b})$ denote the 
inverse matrix of $V_k({\bm a},{\bm b})$ and the $2\times2$ identity matrix, respectively.
Throughout this paper, matrix indices start from zero.
The detailed derivation is provided in Appendix~\ref{sec:single_k_timeevo}.
Here, we take the branch of the square root such that $0\leq {\rm arg}\,\eta_k<\pi$.
Equation~\eqref{eq:evplusmain} implies that the contribution from each $k$ to the eigenvalues is purely imaginary for small $\Gamma$ and becomes real for large $\Gamma$.
This behavior is empirically known for the Liouvillian spectrum in the original SYK model with quenched disorder \cite{kulkarni2022prb} and is similar to that in some non-Hermitian systems \cite{hamazakiprl}.
As shown later, when the values of $\eta_k$ in Eq.~\eqref{eq:evplusmain} are real for all $k$, the corresponding Liouvillian eigenvalues are negative and real, implying non-oscillatory dynamics.

By applying this expression iteratively, we obtain
\begin{align}
    \biggl[\prod_{k=1}^{N/2}\Lambda_k(t)\biggr] \biggl[\prod_{k=1}^{N/2}\langle a_{k} b_{k}\rangle\biggr] 
    =&\sum_{\bm c, \bm d}
    \prod_{k=1}^{N/2} [V_k^{a_k}({\bm a},{\bm b})]_{c_k d_k}[V_k^{-{a_k}}({\bm a},{\bm b})]_{d_k b_k} 
    \exp\biggl[\sum_{k=1}^{N/2}a_k(-1)^{d_k}\eta_k({\bm a},{\bm b}) t\biggr]  
    \nonumber \\
    &\times\exp(\frac{\imi}{2}\sum_{k=1}^{N/2} a_k \sigma_k \epsilon_k H_2 t) \,
    \prod_{k=1}^{N/2} 
    \langle a_{k} c_{k}\rangle  \, \exp(-\frac{\imi}{2}\sum_{k=1}^{N/2} a_k \sigma_k \epsilon_k H_2 t).
\end{align}
We expand the initial density matrix using our CONS as
\begin{align}
    \rho(0)=\sum_{{\bm a},{\bm b}} F({\bm a},{\bm b})\prod_{k} \langle a_k b_k \rangle,
\end{align}
with $F({\bm a},{\bm b})$ being a $c$-number coefficient.
The coefficients satisfy the normalization condition
\begin{equation}
    \sum_{ \bm b}F(\bm a_1,\bm b)=1,
\end{equation}
with $\bm{a}_1=(1,\dots,1)$, as well as positivity and Hermiticity.
The time evolution of the density matrix is given by
\begin{align}
    \rho(t)=&e^{-N \Gamma t}e^{-\imi H_4 t} \left[ \prod_k \Lambda_k(t) \right][\rho(0)]e^{\imi H_4 t} \nonumber \\
    =&e^{-N\Gamma t}\sum_{\bm a, \bm b,\bm c,\bm d}
    F({\bm a},{\bm b})  \exp\left[\sum_k a_k (-1)^{d_k} \eta_k({\bm a},{\bm b})t\right] \prod_k [V_k^{a_k}({\bm a},{\bm b})]_{c_k d_k} [V_k^{-a_k}({\bm a},{\bm b})]_{d_k b_k} \langle a_k c_k \rangle, \label{eq:rho_t_product}
\end{align}
where we have used the following identity
\begin{align}
    \bigg[H_4 -\frac{1}{2}\sum_{k=1}^{N/2} a_k \sigma_k \epsilon_k H_2,\quad \prod_{k=1}^{N/2} \langle a_k b_k \rangle \bigg]=0 \quad \text{for}\, \forall \bm{a},\bm{b} \in \{0,1\}^{N/2}
    \label{eq:h4h4bar},
\end{align}
which is shown in Appendix~\ref{sec:h4h4bar}.
We will show here that we can replace $\eta_k({\bm a},{\bm b})$ with $\eta_k({\bm a},{\bm d})$ in the time-dependent exponential factor in Eq.~\eqref{eq:rho_t_product}.
We fix $k_*$ and consider the replacement $b_{k_*}$ by $ d_{k_*}$ in $\eta_k({\bm a},{\bm b})$. 
We divide the summation into $k=k_*$ and $k\neq k_*$ parts. The first part, $\exp(a_{k_*} (-1)^{d_{k_*}}\eta_{k_*}({\bm a}, {\bm b})t)$, does not depend on $b_{k_*}$ according to Eq.~\eqref{eq:alpha}.
Next, we consider the second part, $\exp(\sum_{k\neq k_*}a_{k} (-1)^{d_{k}}\eta_{k}({\bm a}, {\bm b})t)$.
If $a_{k_*}=1$, $\eta_k({\bm a},{\bm b})$ does not depend on $b_{k_*}$.
Otherwise, i.e., $a_{k_*}=0$, we have $[V_{k_*}^{a_{k_*}}]_{c_{k_*}d_{k_*}}[V_{k_*}^{-a_{k_*}}]_{d_{k_*}b_{k_*}}=\delta_{c_{k_*}d_{k_*}}\delta_{d_{k_*}b_{k_*}}$, which forces 
$d_{k_*}=b_{k_*}$.
In each case, therefore, we may replace $b_{k_*}$ by $ d_{k_*}$.
Since $k_*$ is chosen arbitrarily, $\bm b$ can be replaced by $\bm d$ in the exponential factor.
Hence, we obtain
\begin{align}
     \rho(t)=& \sum_{\bm a,\bm d}
    \exp[t(-N\Gamma +\sum_{k} a_{k} (-1)^{d_{k}} \eta_{k}({\bm a},{\bm d}))]
    \cdot \sum_{\bm b,\bm c} \prod_{k=1}^{N/2} \biggl[[V_k^{a_{k}}({\bm a,\bm b})]_{c_{k},d_{k}}  [V_{k}^{-a_{k}}({\bm a,\bm b})]_{d_{k }b_{k}} \biggr] \nonumber \\
    &\times{\rm Tr} \biggl[ \biggl(\prod_{k=1}^{N/2}\langle a_k b_k \rangle \biggr)^\dagger \rho(0)\biggr] 
    \prod_{k=1}^{N/2} \langle a_k c_k\rangle,
\end{align}
where we have used the following identity,
\begin{align}
    F({\bm a},{\bm b})={\rm Tr} \biggl[ \biggl(\prod_{k=1}^{N/2}\langle a_k b_k \rangle  \biggr)^\dagger \rho(0)\biggr] .
\end{align}
This expression for $\rho(t)$ leads to the eigenvalues of the Liouvillian,
\begin{align}
    \lambda_{{\bm a};{\bm d}}=-N\Gamma  +\sum_{k=1}^{N/2} a_{k} (-1)^{d_{k}} \eta_{k}({\bm a},{\bm d}),
    \label{eq:evlambda}
\end{align}
with the corresponding projection superoperator given by
\begin{align}
    \Pi_{\bm a;\bm d}[\,\bullet\,] 
    =&\sum_{\bm b, \bm c}
    \prod_{k=1}^{N/2} \biggl[[V_k^{a_{k}}({\bm a,\bm b})]_{c_{k},d_{k}}  [V_{k}^{-a_{k}}({\bm a,\bm b})]_{d_{k }b_{k}} \biggr] \cdot {\rm Tr} \biggl[ \biggl(\prod_{k=1}^{N/2}\langle a_k b_k \rangle \biggr)^\dagger \bullet \, \biggr] 
    \prod_{k=1}^{N/2} \langle a_kc_k\rangle.
    \label{eq:projection}
\end{align}
\end{widetext}
The proof of the orthogonality and completeness of this decomposition is provided in Appendix~\ref{sec:orthogonality}.
Note that the present derivation relies on the 
diagonalizability of Eq.~\eqref{eq:vmatrix}
and therefore does not apply exactly at exceptional points.
Nevertheless, exceptional points can influence the dynamics in their vicinity, as discussed for the SYK model \cite{Zheng2026prd}.

\begin{figure*}
    \includegraphics[width=\linewidth]{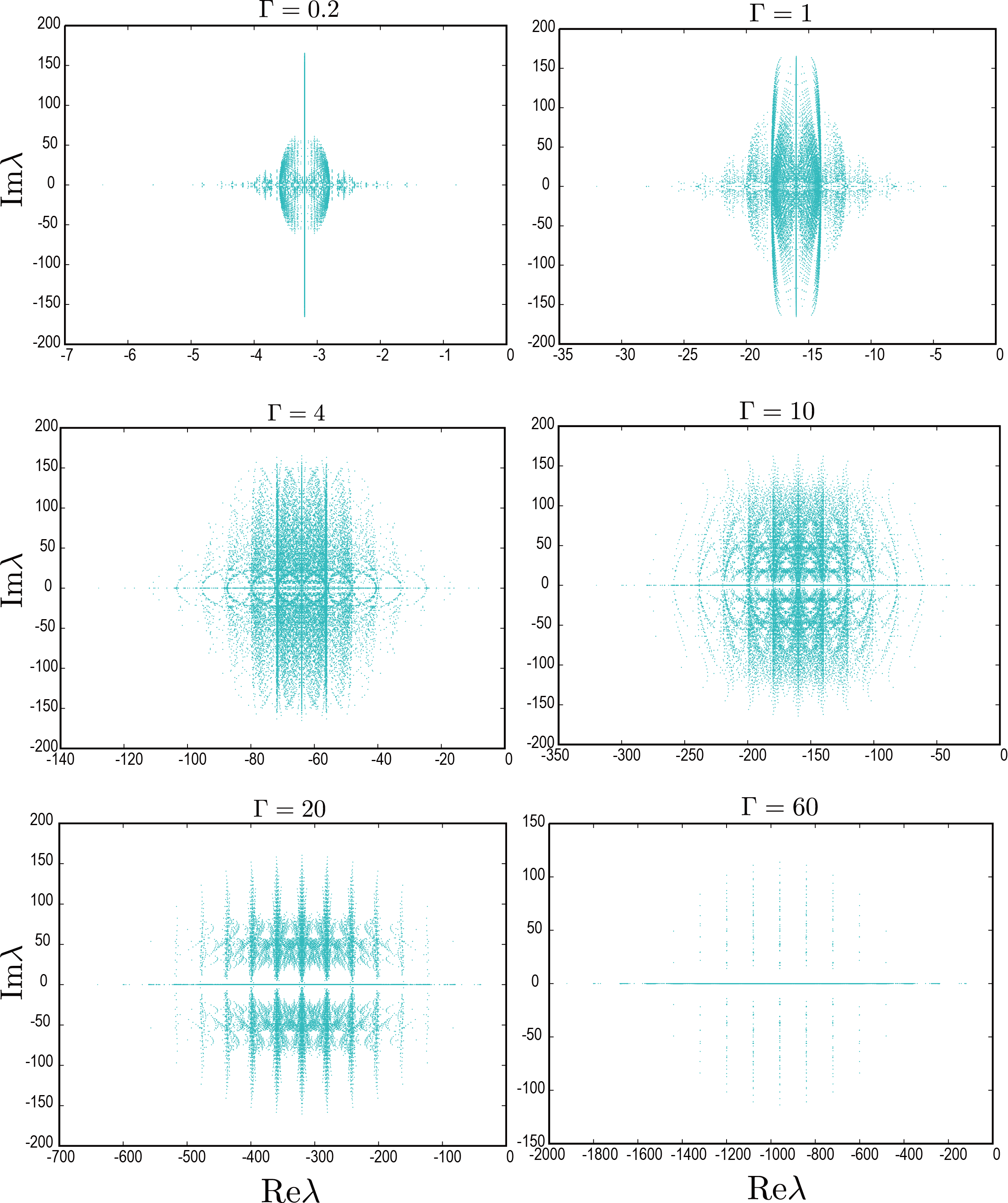}
    \caption{Eigenvalues of the Liouvillian $\mathcal{L}$ 
    for $N=16$ at $\Gamma=0.2, 1, 4, 10, 20$, and $60$.}
    \label{fig:ev}
\end{figure*}

The Liouvillian eigenvalues obtained for $N=16$ are shown in Fig.~\ref{fig:ev}. 
At small $\Gamma$ $(\sim 0.1)$, a high density of eigenvalues is observed along the line ${\rm Re}\,\lambda = -N\Gamma$. 
This is consistent with Eq.~\eqref{eq:evlambda}, where $\eta_k$ is almost purely imaginary [Eq.~\eqref{eq:evplusmain}]. 
As $\Gamma$ increases, pairs of these eigenvalues sequentially collide and subsequently spread along the real axis. 
This process leads to the formation of the stripe structures observed in the intermediate regime ($4 \lesssim \Gamma \lesssim 20$). 
For large $\Gamma$ $(\sim 60)$, most of the eigenvalues become negative real values. 
This is because $\eta_k$ becomes real for sufficiently large $\Gamma$ as the argument of the square root becomes positive, and the imaginary contributions to $\lambda$ vanish due to the cancellation of the time-dependent phase factor as shown in Eq.~\eqref{eq:rho_t_product}.

\begin{figure}
    \includegraphics[width=\linewidth]{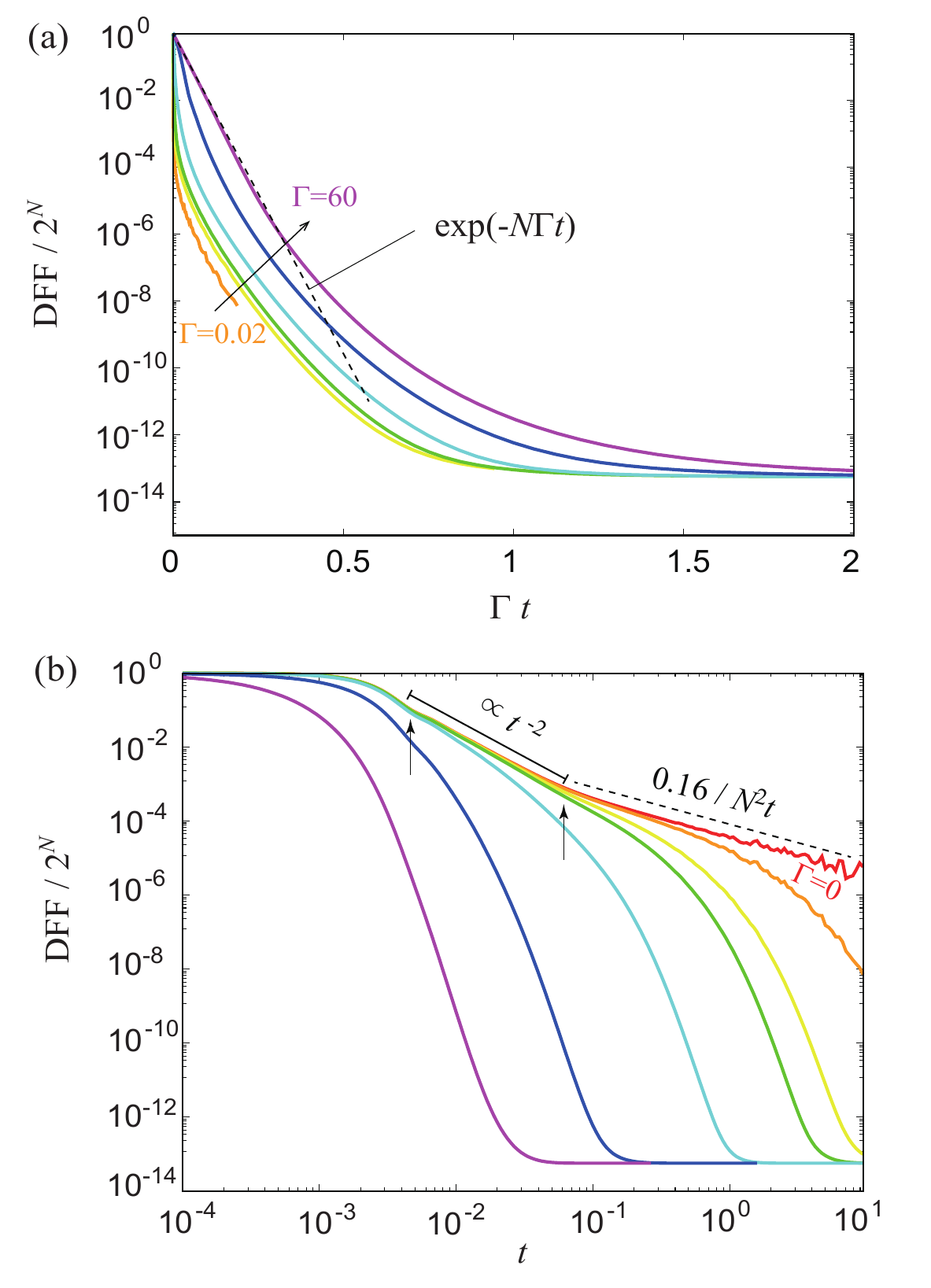}
    \caption{Dissipative form factor at $N=44$ for $\Gamma= 0.02, 0.1, 0.2, 1.0, 10.0, 60.0$ as functions of (a) $\Gamma t$ and (b) $t$. In (a), the black dashed line represents an exponential decay $\exp(-N\Gamma t)$. In (b), the $\Gamma=0$ case is also shown in red for comparison, which corresponds to the spectral form factor (SFF) of $H_4$ at $T=\infty$. The black solid line indicates the $t^{-2}$ scaling, and the black dashed line shows the asymptotic form of the SFF \cite{ozaki-katsura}. Changes of power observed in the $\Gamma=0$ curve are indicated by the arrows.}
    \label{fig:dff}
\end{figure}

Finally, we compute the DFF defined as
\begin{align}
    \textrm{DFF}(t)&=|{\rm Tr} \,e^{t\mathcal{L}}|
    =\left|\sum_{\bm a, \bm d}  \exp\bigl(\lambda_{{\bm a}; {\bm d}}t\bigr)\right|,
\end{align}
where the trace is taken within the operator space. 
\soo{
Figure~\ref{fig:dff} shows the DFF normalized by $2^N$ for several values of $\Gamma$, plotted as a function of (a) $\Gamma t$ on a linear scale and (b) $t$ on a logarithmic scale for $N=44$.
Note that this size is not accessible by straightforward numerical diagonalization.
As shown in Fig.~\ref{fig:dff}(a), for $\Gamma=60$, the DFF exhibits an early-time exponential decay, described by $\exp (-N\Gamma t)$, whereas such behavior is not observed for smaller $\Gamma$. At later times, the DFF rapidly 
approaches its saturation value on a time scale of $1/\Gamma$ for all $\Gamma$.
The unnormalized saturation value is 1; because the plotted quantity is normalized by $2^N$, 
this appears in the figure as $2^{-44}\simeq5.68\times 10^{-14}$.
}

As shown in Fig.~\ref{fig:dff}(b), a logarithmic scale reveals more complex behavior for smaller $\Gamma$.
In the dissipationless limit ($\Gamma\to 0$), the DFF approaches the spectral form factor (SFF) at infinite temperature, defined as $g(t, T=\infty)=|{\rm Tr}\,e^{\imi H_4 t}|^2/2^{N}$, up to a normalization factor. In the SFF, we observe three distinct time regimes characterized by their scaling behavior: constant, $t^{-2}$, and $t^{-1}$. 
The crossover points between these regimes are indicated by arrows.
For the $t^{-1}$ regime, the functional form was previously reported as $0.16/N^2t$ for sufficiently large $N$ \cite{ozaki-katsura}. 

We find that, as $\Gamma\to 0$, the DFF converges to the SFF at infinite temperature in a stepwise manner, reflecting the underlying time regimes.
In the earliest regime, the DFF already agrees with the SFF for $\Gamma \lesssim 1.0$. 
In the intermediate regime, convergence requires a smaller value, $\Gamma\lesssim 0.2$.
As seen from these results, the scale of $\Gamma$ required for convergence to the SFF depends on the time regime.
This behavior is qualitatively different from that of the DFF in the original SYK model \cite{kawabata2023prb}.

\section{Examples\label{sec:examples}}

We consider the time evolution of the density matrix for several initial conditions at $N=4$.

\subsubsection{$\rho_1=\langle11\rangle \langle 11 \rangle(=\bar{f}_1^\dagger\bar{f}_1 \bar{f}_2^\dagger\bar{f}_2)$}
The projected density matrix is given by
\begin{align}
    \Pi_{11;d_1 d_2}[\rho_1]
    =&\sum_{c_1,c_2=0,1}[V]_{c_1 d_1}[V^{-1}]_{d_1 1}[V]_{c_2 d_2}[V^{-1}]_{d_2 1} \nonumber \\
    &\times \langle 1c_1\rangle  \langle 1 c_2\rangle,
\end{align}
with
\begin{equation}
    V=2\Gamma\begin{pmatrix}
        1 & -1 \\
        1 & 1
    \end{pmatrix}.
\end{equation}
Straightforward calculations lead to
\begin{align}
    \Pi_{11;00}[\rho_1]=
    & \frac{1}{4}( \langle 10 \rangle \langle10\rangle+ \langle 10 \rangle \langle11\rangle
    +\langle 11 \rangle \langle10\rangle+ \langle 11 \rangle \langle11\rangle) 
    \nonumber \\
    =&\frac{I}{4}, \\
    \Pi_{11;01}[\rho_1]=
    & \frac{1}{4}( -\langle 10 \rangle \langle10\rangle+ \langle 10 \rangle \langle11\rangle
    -\langle 11 \rangle \langle10\rangle+ \langle 11 \rangle \langle11\rangle),
    \\
    \Pi_{11;10}[\rho_1]=
    & \frac{1}{4}( -\langle 10 \rangle \langle10\rangle - \langle 10 \rangle \langle11\rangle
    +\langle 11 \rangle \langle10\rangle+ \langle 11 \rangle \langle11\rangle),\\
    \Pi_{11;11}[\rho_1]=
    & \frac{1}{4}( \langle 10 \rangle \langle10\rangle - \langle 10 \rangle \langle11\rangle
    -\langle 11 \rangle \langle10\rangle+ \langle 11 \rangle \langle11\rangle).
\end{align}
The completeness condition 
\begin{equation}
    \sum_{d_1,d_2}\Pi_{11;d_1 d_2}[\rho_1]=\rho_1
\end{equation} is easily verified.
These projections lead to the spectral decomposition,
\begin{equation}
    \mathcal{L}\Pi_{11,d_1 d_2}[\rho_1]=-4(d_1+d_2)\Gamma \Pi_{11,d_1 d_2}[\rho_1].
\end{equation}
This result confirms that the state $\Pi_{11;00}[\rho_1]=I/4$ is the steady state as we discussed in Sec.~\ref{sec:spectraldecomposition}.
We note that all the eigenvalues of the Liouvillian 
are real in this case.

\subsubsection{$\rho_2=I/4+\varepsilon (\bar{f}_1+\bar{f}_1^\dagger)\bar{f}_2 \bar{f}_2^\dagger$}
Next, we consider $\rho_2=I/4+\so{\varepsilon}(\bar{f}_1 + \bar{f}^\dagger_1)\bar{f}_2\bar{f}_2^\dagger$, with $|\varepsilon|\leq \frac{1}{4}$
ensuring positive semidefiniteness.
The projected components are
\begin{align}
    \Pi_{11;00}[\rho_2]&= \frac{1}{4} \sum_{b_1,b_2,c_1,c_2} [V]_{c_1 0} [V^{-1}]_{0 b_1} [V]_{c_20} [V^{-1}]_{0 b_2} \langle1 c_1 \rangle \langle1 c_2\rangle \nonumber \\
    &=\frac{1}{4}I,\\
    \Pi_{01;0 d_2}[\rho_2]&= \so{\varepsilon} \sum_{c_2=0,1}[V_{2-}]_{c_2 d_2}[V_{2-}^{-1}]_{d_20} \langle 00 \rangle \langle 1c_2 \rangle, \\
    \Pi_{01;1 d_2}[\rho_2]&= \so{\varepsilon} \sum_{c_2=0,1}[V_{2+}]_{c_2 d_2}[V_{2+}^{-1}]_{d_20} \langle 01 \rangle \langle 1c_2 \rangle, 
\end{align}
where 
\begin{align}
    V_{2\pm}=\begin{pmatrix}
        \pm 2\imi +2\sqrt{\Gamma^2-1} & -2\Gamma \\
        2\Gamma &    \pm 2\imi +2\sqrt{\Gamma^2-1}
    \end{pmatrix},
\end{align}
and we have used $\epsilon_1 \epsilon_2=4$ for $N=4$.
These components satisfy the completeness relation,
\begin{equation}
    \Pi_{11;00}[\rho_2]+ \sum_{d_1,d_2} \Pi_{01;d_1 d_2}[\rho_2]=\rho_2.
\end{equation}
As in the previous example, $\Pi_{11;00}$ is the projection with eigenvalue 0,
\begin{equation}
    \mathcal{L}\Pi_{11;00}[\rho_2]=0,
\end{equation}
indicating the steady state.
The remaining projected components are given by 
\begin{widetext}
\begin{align}
    &\Pi_{01;00}[\rho_2] = \frac{\so{\varepsilon}}{D_-} \biggl[  (-2\imi + 2\sqrt{\Gamma^2-1})^2 \langle 00 \rangle \langle 10 \rangle +2\Gamma (-2\imi + 2\sqrt{\Gamma^2-1}) \langle 00\rangle \langle 11\rangle  \biggr],\\
    &\Pi_{01;01}[\rho_2] = \frac{\so{\varepsilon}}{D_-} \biggl[  4\Gamma^2 \langle 00 \rangle \langle 10 \rangle -2\Gamma (-2\imi + 2\sqrt{\Gamma^2-1}) \langle 00\rangle \langle 11\rangle  \biggr],\\
    &\Pi_{01;10}[\rho_2] = \frac{\so{\varepsilon}}{D_+} \biggl[  (2\imi + 2\sqrt{\Gamma^2-1})^2 \langle 01 \rangle \langle 10 \rangle +2\Gamma (2\imi + 2\sqrt{\Gamma^2-1}) \langle 01\rangle \langle 11\rangle  \biggr],\\
    &\Pi_{01;11}[\rho_2] = \frac{\so{\varepsilon}}{D_+} \biggl[  4\Gamma^2 \langle 01 \rangle \langle 10 \rangle -2\Gamma (2\imi + 2\sqrt{\Gamma^2-1}) \langle 01\rangle \langle 11\rangle  \biggr],
\end{align}
\end{widetext}
where $D_\pm=\bigl(\pm2\imi +2\sqrt{\Gamma^2-1}\bigr)^2 + 4\Gamma^2$.

Equation~\eqref{eq:evlambda} gives
\begin{equation}
\mathcal{L}\Pi_{01;d_1d_2}=\left(-4\Gamma+2\cdot(-1)^{d_2} \sqrt{\Gamma^2-1}\right)\Pi_{01;d_1d_2}[\rho_2].
\end{equation}
This can be confirmed by directly applying $\mathcal{L}$ to the relevant operator-basis elements:
\begin{align}
    \begin{pmatrix}
        \mathcal{L}[\langle00\rangle \langle10\rangle] \\
        \mathcal{L}[\langle00\rangle \langle11\rangle]
    \end{pmatrix}
    &=
    \begin{pmatrix}
        -2\imi-4\Gamma & 2\Gamma \\
        2\Gamma & 2\imi -4\Gamma
    \end{pmatrix}
    \begin{pmatrix}
        \langle 00 \rangle \langle 10 \rangle \\
        \langle 00 \rangle \langle 11 \rangle
    \end{pmatrix}, \\
    \begin{pmatrix}
        \mathcal{L}[\langle01\rangle \langle10\rangle] \\
        \mathcal{L}[\langle01\rangle \langle11\rangle]
    \end{pmatrix}
    &=
    \begin{pmatrix}
        2\imi-4\Gamma & 2\Gamma \\
        2\Gamma & -2\imi -4\Gamma
    \end{pmatrix}
    \begin{pmatrix}
        \langle 01 \rangle \langle 10 \rangle \\
        \langle 01 \rangle \langle 11 \rangle
    \end{pmatrix}.
\end{align}

This result implies the existence of an exceptional point at $\Gamma=1$; the eigenvalues of the Liouvillian are complex for $\Gamma<1$ and real and negative for $\Gamma>1$.
This behavior is reminiscent of the case of the original SYK model with dissipation \cite{kulkarni2022prb}.

\twocolumngrid

\section{Two-point correlation function\label{sec:correlation} }
In this section, we study the steady-state two-point correlation function defined as
\begin{equation}
    C^{i,j}_{\infty}(t)=
    {\rm Tr}\left[\rho_\infty \gamma_i(t) \gamma_j\right]=
    \frac{1}{2^{N/2}}
    {\rm Tr} [e^{t \mathcal{L}^\dagger}[\gamma_i] \gamma_j]
    \label{eq:two_point},
\end{equation}
where $\mathcal{L}^\dagger$ is the adjoint of the Liouvillian, which describes the time evolution of an operator $A$ as follows:
\begin{equation}
  \frac{d A}{d t}=\mathcal{L}^\dagger [A]=\imi[H, A]+\sum_{j=1}^N \left(L_{j}^{\dagger} A L_{j}-\frac{1}{2}\left\{L_{j}^{\dagger} L_{j}, A\right\}\right).
  \label{eq:Lindblad_adjoint}
\end{equation}
In our case, $H=H_4$ and $L_j=\sqrt{\Gamma}\gamma_j$. Then, we can rewrite Eq.~\eqref{eq:Lindblad_adjoint} as
\begin{equation}
  \mathcal{L}^\dagger [A]=\imi[H_4, A]+2 \Gamma\sum_{k=1}^{N/2} \left(f_{k}^\dagger A f_{k}+f_{k} A f_{k}^\dagger-A\right).
\end{equation}
The eigenvalues and 
corresponding projection superoperators
of $\mathcal{L}^\dagger$ can be calculated in the same way as those of $\mathcal{L}$.
\begin{align}
    \lambda_{{\bm a};{\bm d}}=-N\Gamma  +\sum_{k=1}^{N/2} a_{k} (-1)^{d_{k}} \eta_{k}({\bm a},{\bm d}),
\end{align}

\begin{align}
    \widetilde{\Pi}_{\bm a;\bm d}[\,\bullet\,] 
    =&\sum_{\bm b, \bm c}
    \prod_{k=1}^{N/2} \biggl[[\widetilde{V}_k^{a_{k}}({\bm a,\bm b})]_{c_{k},d_{k}}  [\widetilde{V}_{k}^{-a_{k}}({\bm a,\bm b})]_{d_{k }b_{k}} \biggr] \nonumber \\
    &\times{\rm Tr} \biggl[ \biggl(\prod_{k=1}^{N/2}\langle a_k b_k \rangle \biggr)^\dagger \bullet \, \biggr] 
    \prod_{k=1}^{N/2} \langle a_kc_k\rangle.
    \label{eq:eigenprojection}
\end{align}
where
\begin{align}
    \widetilde{V}_k({\bm a},{\bm b})=\begin{pmatrix}
        -\frac{\imi}{2} \alpha_k({\bm a},{\bm b}) + \eta_k({\bm a},{\bm b}) & -2\Gamma \\
        2\Gamma & -\frac{\imi}{2} \alpha_k({\bm a},{\bm b})+\eta_k({\bm a},{\bm b}) 
    \end{pmatrix}.
\end{align}
To calculate Eq.~\eqref{eq:two_point}, we first consider $\widetilde{\Pi}_{\bm a;\bm d}[f_l]$.
Noting that  
$f_l=\bar{f}_l\prod_{j=1}^{N/2} (\bar{f}^\dagger_j \bar{f}_j-\bar{f}_j\bar{f}^\dagger_j)$, we only need to consider the sectors where
\begin{align}
    a_l=0
\end{align}
and
\begin{equation}
    a_k= 1\quad\forall k\neq l.
\end{equation}
In what follows, expressing this sector by writing the subscript $l$ instead of $\bm{a}$, and defining $G_{k,l}=\sqrt{4\Gamma^2-\frac{(\epsilon_k \epsilon_l)^2}{4}}$, we have
\begin{equation}
    \lambda_{l,\bm{b}} =-N\Gamma  +\sum_{k\neq l} (-1)^{b_{k}} G_{k,l},
    \label{eq:two_point_spectrum}
\end{equation}
and
\begin{align}
    \widetilde{\Pi}_{l,\bm{b}}[f_l]&=f_l
    \prod_{k\neq l}\left(\left[U_{k,l}^{-1}\right]_{b_{k},0}-\left[U_{k,l}^{-1}\right]_{b_{k},1}
    \right)\nonumber\\
    &\times \left([U_{k,l}]_{0,b_{k}} f_k f_k^\dagger
    -[U_{k,l}]_{1,b_{k}} f_k^\dagger f_k\right),
\end{align}
where 
\begin{align}
    U_{k,l}=\begin{pmatrix}
        \imi \frac{\epsilon_k \epsilon_l}{2} + G_{k,l}& -2\Gamma \\
        2\Gamma & \imi \frac{\epsilon_k \epsilon_l}{2}+G_{k,l}
    \end{pmatrix}.
\end{align}
Using these 
projection superoperators, we obtain a closed form for Eq.~\eqref{eq:two_point}:

\begin{widetext}
    \begin{align}
    C^{i,j}_{\infty}(t)
    =\frac{2e^{-N\Gamma t}}{ N}\sum_{k=1}^{N/2}\Biggl(\cos\left[(i-j) \theta_k\right]
    \prod_{n\neq k} \left[\cosh(G_{n,k}t)
-\frac{2\Gamma}{G_{n,k}}\sinh(G_{n,k}t)\right]\Biggr).
    \label{eq:autocorrelation}
\end{align}
\end{widetext}

\begin{figure}
 \centering
  \includegraphics[width=\linewidth]{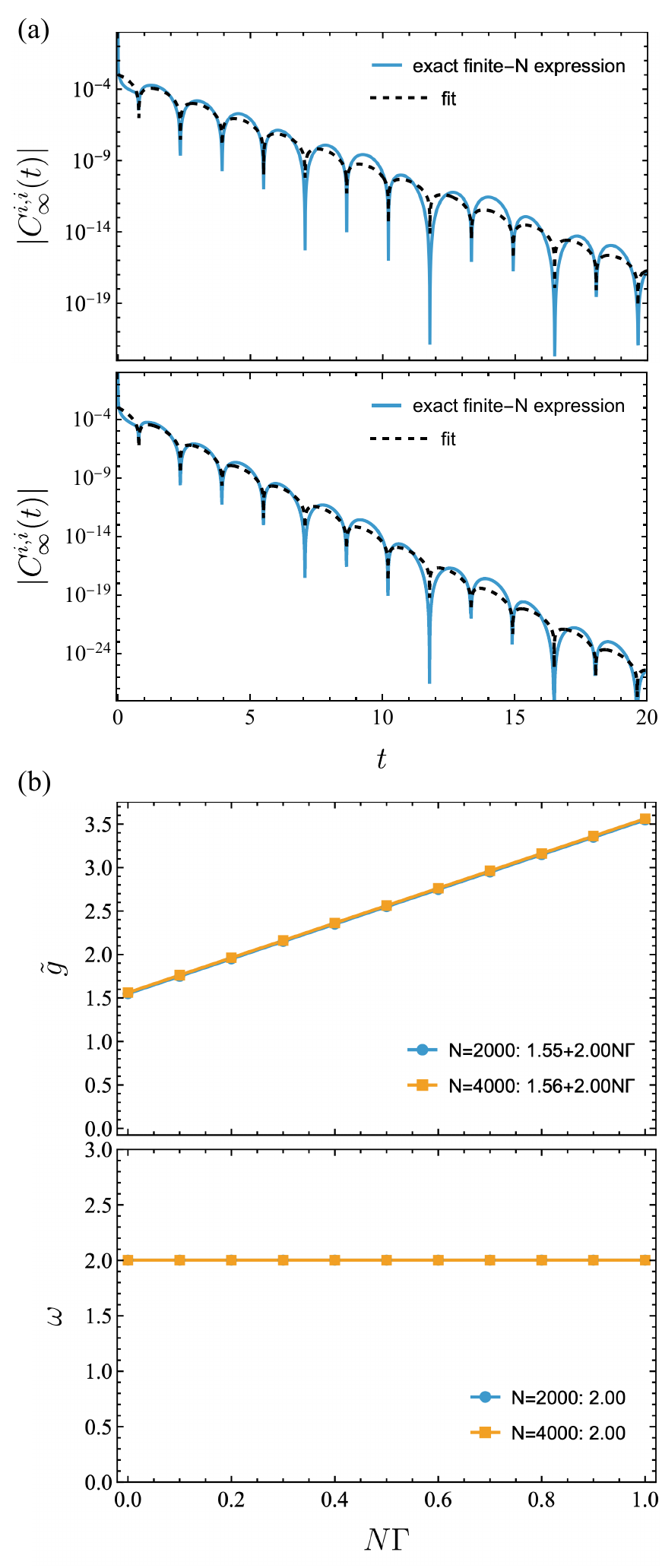}
  \caption{(a) Steady-state autocorrelation function for \(N=4000\), shown for \(N\Gamma=0\) (top) and \(N\Gamma=0.5\) (bottom). The blue line is computed using Eq.~\eqref{eq:autocorrelation}, and the dotted lines represent fits $Ae^{-\tilde{g} t}|\cos(\omega t+\phi)|$ using data points from $t=1$ to $t=20$. (b) Values of $\tilde{g}$ and $\omega$ obtained by fitting the autocorrelation function for different $\Gamma$. Data for $N =2000$ and $4000$ are shown.
}
  \label{fig:correlation}
\end{figure}

In the following, we focus on the case where $i=j$, that is, the steady-state autocorrelation function $C^{i,i}_{\infty}(t)$. With $N\Gamma:=\xi$ and $t>0$ fixed, its large-$N$ limit can be evaluated as follows (see Appendix~\ref{sec:correlation_analytics} for the derivation):\begin{equation}
C^{i,i}_{\infty}(t)
=
\frac{2e^{-2 \xi t}}{N}
\sum_{m=0}^{\infty}
\Phi[(2m+1)t]
+o(N^{-1})
\label{eq:analytics}
\end{equation}
where 
\begin{equation}
    \Phi(u)=\prod_{r=0}^{\infty}
\cos \left( \frac{2u}{2r+1} \right)=\cos(2u)\prod_{r=1}^{\infty}
\cos \left( \frac{2u}{2r+1} \right).
\end{equation}
Note that Eq.~\eqref{eq:analytics} is valid for $\Gamma=0$, but not at $t=0$. Indeed, since $\Phi(0)=1$, the right-hand side of Eq.\eqref{eq:analytics} diverges at $t=0$. Also, since $\Phi(u)=0$ for $u=\frac{\pi}{4}(2j+1)$, $j=0,1,2,\ldots$, all terms $\Phi[(2m+1)t]$ vanish simultaneously when
\begin{equation}
    t=\frac{\pi}{4}(2j+1), \quad j=0,1,2,\ldots .
\end{equation}
Therefore, $C^{i,i}_{\infty}(t)$ also vanishes at these times. For large $u$, its envelope decays exponentially (see Appendix~\ref{sec:correlation_analytics}),
\begin{equation}
|\Phi(u)|\simeq \exp\!\left(-\frac{\pi}{2}u\right)\cos(2u)
\label{eq:phi_asymptotics}
\end{equation}
up to oscillatory factors. Therefore, the long-time decay of the steady-state autocorrelation function is given by:
\begin{equation}
    |C^{i,i}_{\infty}(t)|\simeq\exp\!\left[-\left(\frac{\pi}{2}+2\xi\right) t\right]\cos(2t).
\end{equation}
In Fig.~\ref{fig:correlation}, we plot Eq.~\eqref{eq:analytics} at finite $N$ to verify this behavior. The blue curve in Fig.~\ref{fig:correlation}(a) shows the result for $N=4000$ and $\Gamma=0$. We observe zeros at $t=\frac{\pi}{4}(2j+1)$ $j=0,1,2,\ldots$, together with exponential decay at long times. The dotted curve is a fit $Ae^{-\tilde{g} t}|\cos(\omega t+\phi)|$ to the blue curve, which agrees well with the data except 
at very short times. Figure~\ref{fig:correlation}(b) shows the fitted coefficients $\tilde{g}$ and $\omega$ as functions of $N\Gamma$. For both $N=2000$ and $4000$, we find behavior consistent with $\tilde{g}=\pi/2+2\xi$ and $\omega=2$, in agreement with the large-$N$ result.

It is worth noting that the autocorrelation function exhibits exponential decay even at $\Gamma=0$. Remarkably, the origin of this decay is already encoded in the spectrum of the Liouvillian. Since only eigenvalues of the form Eq.~\eqref{eq:two_point_spectrum} contribute to the two-point correlation function, we define the corresponding asymptotic decay rate as
\begin{equation}
g=-\max \operatorname{Re}\lambda_{l,\bm{b}}.
\label{eq:asymptotic_decay}
\end{equation}
With this definition, one finds that the limits $N\to\infty$ and $\Gamma\to0$ do not commute:
\begin{equation}
    \lim_{N\to\infty}\lim_{\Gamma\to0}g\neq \lim_{\Gamma\to 0}\lim_{N\to\infty}g.
    \label{eq:noncommute}
\end{equation}
Indeed, taking $\Gamma\to0$ at fixed $N$ gives
\begin{equation}
 \lim_{N\to\infty}\lim_{\Gamma\to0}g=0,
 \label{eq:noncommute1}
\end{equation}
whereas taking the thermodynamic limit first gives
\begin{equation}
   \lim_{\Gamma\to 0}\lim_{N\to\infty}g=\frac{\pi}{2}.
    \label{eq:noncommute2}
\end{equation}
The latter value coincides with the decay rate $\tilde{g}$ extracted from the autocorrelation function at $\Gamma=0$. The derivation of Eqs.~\eqref{eq:noncommute1} and \eqref{eq:noncommute2} is given in Appendix~\ref{sec:singularity}. Similar singular behavior of the spectral gap has been discussed in ergodic quantum many-body systems under the name of anomalous relaxation~\cite{sa_lindbladian_2022,garcia_keldysh_2023,garcia-garcia_UniversalityItsLimits_2023,shackleton_ExactlySolvableDissipative_2024,yoshimura_RobustnessQuantumChaos_2024a,sa_ExactlySolvableDissipative_2025,jacoby_SpectralGapsLocal_2025a,yoshimura_TheoryIrreversibilityQuantum_2025a}. In particular, Ref.~\cite{mori_liouvillian_2024} directly relates this Liouvillian gap singularity to Ruelle--Pollicott resonances.

In the present case, however, the Hamiltonian \(H_4\) is integrable, which makes the appearance of such a singularity in the two-point sector rather unexpected. We believe that this behavior is tied to the rapid growth of the operator size. Indeed, despite the integrability of the clean SYK model, the OTOC exhibits pronounced growth in the absence of dissipation~\cite{ozaki-katsura}. For all-to-all coupled systems in which the average operator size grows exponentially, it has been argued that the operator size can saturate to a system-size-independent value, leading to a \(\Gamma\)-independent decay of the Loschmidt echo~\cite{schuster_OperatorGrowthOpen_2023}. Our results suggest that an analogous mechanism, driven by rapid operator growth and subsequent saturation, may also be at work in the clean SYK model.

\twocolumngrid
\section{Summary \label{sec:summary}}
\so{
We have proposed and studied a clean SYK model with dissipation formulated within the GKSL formalism.
By utilizing the integrability of the clean SYK model together with the peculiar symmetry of the Liouvillian,
we have derived an exact solution in a spectrum-resolved form, i.e., the eigenvalues and corresponding projection superoperators
of the Liouvillian for arbitrary system size $N$.
We have identified transitions in the Liouvillian spectrum from complex to real 
eigenvalues with increasing dissipation strength, associated with exceptional-point structures
ubiquitous in non-Hermitian systems.
The obtained solution allows us to calculate the DFF at $N=44$, which is 
not accessible by straightforward numerical diagonalization.
We find that the DFF converges to the SFF at infinite temperature in the dissipationless limit 
in a stepwise manner.
Such behavior has not been observed in the DFF of the original SYK model with dissipation.
}

\so{
We have determined the scaling of the gap that governs the long-time decay of the 
two-point correlation functions.
Importantly, the gap does not vanish in the dissipationless limit when the thermodynamic limit is taken first.
This phenomenon, known as anomalous relaxation, suggests a possible connection with 
chaotic dynamics and quantum Ruelle-Pollicott resonances.
These findings indicate that the present dissipative clean SYK model 
offers a useful platform for exploring anomalous relaxation and 
other nontrivial open quantum dynamics of many-body systems.
}

Although we focused on the open clean SYK system,
another variant, the clean SUSY SYK model \cite{ozaki-katsura},
has a structure similar to $H_4$, and can be analyzed using the approach developed in this work. 
Furthermore, an integrable clean SYK family with $q$-body interaction has 
recently been identified \cite{fukai-katsura2026}. 
These models may exhibit intriguing open quantum dynamics related to quantum chaos.

\begin{acknowledgments}
We thank Shinsei Ryu and Takato Yoshimura for valuable discussions. H.K. was supported by JSPS KAKENHI Grants No. JP23K25783, No. 23K25790, and MEXT KAKENHI Grant-in-Aid for Transformative Research Areas A “Extreme Universe” (KAKENHI Grant No. JP21H05191). H.Y. was supported by the Special Postdoctoral Researchers Program at RIKEN, Japan Society for the Promotion of Science (JSPS) through the Overseas Research Fellowship, and JSPS KAKENHI Grant No. 26K17059.
\so{The computation in this work was performed using the facilities of the Supercomputer Center, the Institute for Solid State Physics, the University of Tokyo.}
\end{acknowledgments}

\appendix 
\section{Derivation of Eq.~\eqref{eq:single_k_timeevo}}
\label{sec:single_k_timeevo}
To obtain the action of $\Lambda_{k}(t)$, we consider the following initial-value problem for $\bar{\rho}_k(t)$:
\begin{align}
    \frac{d}{dt}\bar{\rho}_k(t)=2\Gamma\mathcal{K}_k(t)[\bar{\rho}_k(t)], \quad \bar{\rho}_k(0)=\rho(0).\label{eq:eomk}
\end{align}
We expand $\bar{\rho}(t)$ as
\begin{align}
    \bar{\rho}_k(t)=\sum_{\bm a,\bm b}F (t; {\bm a},{\bm b}) \prod_{k'=1}^{N/2} \langle a_{k'} b_{k'}\rangle, \label{eq:rhobar} 
\end{align}
where $F(t; {\bm a},{\bm b})$ denotes the expansion coefficients in our CONS. The binary vectors ${\bm a}$ and ${\bm b}$ have components in $\{0,1\}$, and the summation is taken over all such vectors.
First, we consider the action of $\mathcal{K}_k(t)$ on $\prod_{k'=1}^{N/2} \langle a_{k'} b_{k'}\rangle$:
\begin{align}
    \mathcal{K}_k(t)[\prod_{k'=1}^{N/2} \langle a_{k'} b_{k'} \rangle] = &e^{\imi \sigma_k \epsilon_k H_2 t}\prod_{k'=1}^{k-1} \langle a_{k'} b_{k'} \rangle
     \cdot \mathcal{K}_k(0)[\langle a_k b_k \rangle] \nonumber \\
     &\times\prod_{k'=k+1}^{N/2} \langle a_{k'} b_{k'}\rangle
    e^{-\imi \sigma_k \epsilon_k H_2 t}.
    \label{eq:k_action}
\end{align}
Using the identities,
\begin{align}
    &\sigma_k \langle a_k b_k  \rangle = -(-1)^{b_k} \langle a_k b_k \rangle,
    \nonumber \\
    &\langle a_k b_k \rangle \sigma_k = (-1)^{a_k+b_k} \langle a_k b_k \rangle,
    \label{eq:sigmaval}
\end{align}
we find
\begin{align}
    &e^{\imi \sigma_k \epsilon_k H_2 t} \langle a_{k'} b_{k'} \rangle e^{-\imi \sigma_k \epsilon_k H_2 t} \nonumber \\
    &=\exp[-\imi (-1)^{b_{k'}} \sigma_k \epsilon_k \epsilon_{k'}(1-a_{k'})t]  \langle a_{k'} b_{k'} \rangle \quad
    (k\neq k')     
    \label{eq:ab_unitary}
\end{align}
 and 
\begin{align}
    e^{\imi \sigma_k \epsilon_k H_2 t} \mathcal{K}_k(0)[\langle a_{k} b_k  \rangle]e^{-\imi \sigma_k \epsilon_k H_2 t} 
    = \mathcal{K}_k(0) [\langle a_k b_k \rangle].
\end{align}
Inserting the identity operator $I=e^{-\imi \sigma_k \epsilon_k H_2 t}e^{\imi \sigma_k \epsilon_k H_2 t}$ between each neighboring pair of basis in Eq.~\eqref{eq:k_action}, the time-dependent parts can be combined into a single factor, and we obtain
\begin{align}
&\mathcal{K}_k(t) \biggl[\prod_{k'=1}^{N/2} \langle a_{k'} b_{k'} \rangle \biggr] \nonumber \\
    &=e^{\imi \alpha_k({\bm a},{\bm b}) \sigma_k t} 
      \prod_{k'=1}^{k-1} \langle a_{k'} b_{k'} \rangle \cdot
     \mathcal{K}_k(0)\big[\langle a_k b_k \rangle\big] \cdot \prod_{k'=k+1}^{N/2} \langle a_{k'} b_{k'} \rangle.
     \label{eq:kkoperation}
\end{align}

In the following, we abbreviate $F(t;{\bm a},{\bm b})$ as $F_{a_k,b_k}(t)$ and derive their equations of motion.
Substituting Eq.~\eqref{eq:rhobar} into Eq.~\eqref{eq:eomk} and using Eq.~\eqref{eq:kkoperation}, we obtain
\begin{align}
    &\dot{F}_{0,0}=\dot{F}_{0,1}=0,\nonumber \\
    &\dot{F}_{1,0}= 2\Gamma e^{-\imi\alpha_k({\bm a},{\bm b}) t} F_{1,1}, \nonumber \\
    &\dot{F}_{1,1}= 2\Gamma e^{\imi \alpha_k({\bm a}, {\bm b}) t}F_{1,0}.
\end{align}
Note that $\alpha_k({\bm a},{\bm b})$ is independent of both $a_k$ and $b_k$.
In the $a_k=0$ subspace, the equations are solved trivially, giving $F_{0,0}(t)=F_{0,0}(0)$ and $F_{0,1}(t)=F_{0,1}(0)$.
To solve the $a_k=1$ subspace, we set
\begin{align}
    F_{1,0}(t)=e^{-\frac{\imi}{2}\alpha_k({\bm a},{\bm b})t}u_0(t), 
    \quad F_{1,1}(t)=e^{\frac{\imi}{2}\alpha_k({\bm a},{\bm b})t}u_1(t),
\end{align}
which leads to
\begin{align}
    \begin{pmatrix}
        \dot{u}_0 \\ \dot{u}_1
    \end{pmatrix}
    =\begin{pmatrix}
        \frac{\imi}{2} \alpha_k({\bm a},{\bm b}) & 2\Gamma \\
        2\Gamma & -\frac{\imi}{2} \alpha_k({\bm a},{\bm b})
    \end{pmatrix}
    \begin{pmatrix}
        u_0 \\u_1
    \end{pmatrix}
    \label{eq:ueom}.
\end{align}
This structure frequently appears in non-Hermitian systems with $\mathcal{P}\mathcal{T}$ symmetry,
and its exceptional points are $\Gamma=\frac{1}{4} |\alpha_{k}({\bm a},{\bm b})|$.
Except at the exceptional point, the solutions for $u_0(t)$ and $u_1(t)$ are given by
\begin{align}
    \begin{pmatrix}
        u_0(t) \\ u_1(t)
    \end{pmatrix}
    =V_k
    \begin{pmatrix}
        e^{\eta_k t} & 0 \\
        0 & e^{-\eta_k t}
    \end{pmatrix}
    V_k^{-1}
    \begin{pmatrix}
        F_{1,0}(0) \\ F_{1,1}(0)
    \end{pmatrix},
\end{align}
with
\begin{align}
    \eta_k({\bm a},{\bm b}) 
    = \sqrt{4 \Gamma^2-\frac{1}{4}\alpha_k({\bm a},{\bm b})^2}
    \label{eq:evplus}
\end{align}
and 
\begin{align}
    V_k({\bm a},{\bm b})=\begin{pmatrix}
        \frac{\imi}{2} \alpha_k({\bm a},{\bm b}) + \eta_k({\bm a},{\bm b}) & -2\Gamma \\
        2\Gamma & \frac{\imi}{2} \alpha_k({\bm a},{\bm b})+\eta_k({\bm a},{\bm b}) 
    \end{pmatrix}.
\end{align}
These solutions can be expressed in the unified form,
\begin{align}
    \rhofunc_{a_k,b_k}(t)=&\exp[-a_k (-1)^{b_k} \frac{\imi}{2}\alpha_k({\bm a},{\bm b})t] \nonumber \\
    &\times \exp[a_k (-1)^{d_k} \eta_k({\bm a},{\bm b}) t] \nonumber \\
    &\times \sum_{c_k, d_k=0,1} [V_k^{a_k}]_{b_k d_k}  
    [V_k^{-a_k}]_{d_k c_k} \rhofunc_{a_k,c_k}(0).
\end{align}
Substituting the solutions for $F_{a_k,b_k}(t)$ into Eq.~\eqref{eq:rhobar}, we obtain
\begin{widetext}
\begin{align}
    \bar{\rho}_k(t)=&\sum_{\bm a,\bm b} \rhofunc_{a_k, b_k}(t)  \prod_{k'=1}^{N/2} 
    \langle a_{k'} b_{k'} \rangle \nonumber \\
    =&\sum_{\bm a,\bm b}\sum_{c_k,d_k}\exp[a_k (-1)^{d_k}\eta_k t]   [V_k^{a_k}]_{b_k d_k}[V_k^{-a_k}]_{d_k c_k}
    \rhofunc_{a_k c_k}(0)  \exp[-a_k \sigma_k \epsilon_k \frac{\imi t}{2} \sum_{k'\neq k}\epsilon_{k'}(1-a_{k'})(-1)^{b_{k'}}] \prod_{k'=1}^{N/2} \langle a_{k'}b_{k'}\rangle.
    \label{eq:rhobarsol}
\end{align}
Making use of Eq.~\eqref{eq:ab_unitary}, we have
\begin{align}
    &\exp[-a_k \sigma_k \epsilon_k \frac{\imi t}{2} \sum_{k'\neq k}\epsilon_{k'}(1-a_{k'})(-1)^{b_{k'}}]
    \prod_{k'=1}^{N/2} \langle a_{k'}b_{k'}\rangle \nonumber \\
    &=\prod_{k'=1}^{k-1} \exp \big(\frac{\imi}{2} a_k\sigma_k \epsilon_k H_2  t\big)
    \langle a_{k'} b_{k'} \rangle
    \exp\big(-\frac{\imi}{2} a_k\sigma_k \epsilon_k H_2  t\big)
    \cdot \langle a_k b_k \rangle  \cdot \prod_{k'=k+1}^{N/2} \exp \big(\frac{\imi}{2} a_k\sigma_k \epsilon_k H_2  t\big)
    \langle a_{k'} b_{k'} \rangle
    \exp \big(-\frac{\imi}{2} a_k\sigma_k \epsilon_k H_2  t\big) \nonumber \\
    &=\exp \big(\frac{\imi}{2} a_k\sigma_k \epsilon_k H_2  t\big)
    \biggl[\prod_{k'}\langle a_{k'} b_{k'} \rangle \biggr] 
    \exp \big(-\frac{\imi}{2} a_k\sigma_k \epsilon_k H_2  t\big).
    \label{eq:ab_unitary_inv}
\end{align}
From the definition of $\rhofunc_{a_k,c_k}$, we obtain
\begin{align}
    \rhofunc_{a_k, c_k}(0)={\rm Tr}[\{\prod_{k'=1}^{k-1} \langle a_{k'}b_{k'}\rangle \cdot \langle a_k c_k\rangle \cdot \prod_{k'=k+1}^{N/2}\langle a_{k'} b_{k'}\rangle \}^\dagger 
    \rho(0)].
\end{align}

To see how $\mathcal{K}_k(t)$ acts on a single basis element, we take 
the initial state to be $\rho(0)=\prod_{k=1}^{N/2} \langle \tilde{a}_k \tilde{b}_k\rangle$.
Then, the coefficient $F_{a_k, c_k}(0)$ is obtained as
\begin{equation}
    F_{a_k c_k}(0)=\delta_{a_1 \tilde{a}_1} \cdots \delta_{a_{N/2} \tilde{a}_{N/2} } \cdot 
    \delta_{b_1 \tilde{b}_1 } \cdots \delta_{c_k \tilde{b}_k} \cdots \delta_{b_{N/2}\tilde{b}_{N/2} }. \label{eq:F0}
\end{equation}
Combining Eqs.~\eqref{eq:rhobarsol}, \eqref{eq:ab_unitary_inv}, and \eqref{eq:F0} and summing over ${\bm a}$, ${\bm b}$ (except $b_k$), and $c_k$, we obtain
\begin{align}
    \bar{\rho}_k(t) =& \sum_{b_k,d_k} \exp[\tilde{a}_k (-1)^{d_k} \eta_k(\tilde {\bm a}, \tilde{\bm b}')t] [V_k^{{\tilde a}_k}(\tilde {\bm a}, \tilde{\bm b}')]_{b_k d_k} [V_k^{-{\tilde a}_k}(\tilde {\bm a}, \tilde{\bm b}')]_{d_k \tilde{b}_k} 
    \nonumber \\
    & \times \exp \big(\frac{\imi}{2} \tilde{a}_k\sigma_k \epsilon_k H_2  t\big)
    \biggl[\prod_{k'=1}^{k-1}\langle \tilde{a}_{k'} \tilde{b}_{k'} \rangle 
    \cdot \langle \tilde{a}_{k}b_k\rangle \cdot 
    \prod_{k'=k+1}^{N/2}\langle \tilde{a}_{k'} \tilde{b}_{k'} \rangle 
    \biggr] 
    \exp \big(-\frac{\imi}{2} \tilde{a}_k\sigma_k \epsilon_k H_2  t\big),
\end{align}
with ${\tilde{\bm b}}'=(\tilde{b}_1, \dots,b_k,\dots,\tilde{b}_{N/2})$.
Since $\alpha_k(\tilde{\bm a},\tilde{\bm b})$ and $\eta_k(\tilde{\bm a},\tilde{\bm b})$ do not depend on $\tilde{b}_k$, we can substitute $\tilde{\bm b}' \to \tilde{\bm b}$ in $\eta_k(\tilde {\bm a}, \tilde{\bm b}')$ appearing in the exponential factor and in $V_k^{\pm{\tilde a}_k}(\tilde{\bm a}, \tilde{\bm b}')$.
We first relabel $b_k$ as $c_k$ and then drop the tildes, $(\tilde{\bm a},\tilde{\bm b})\to (\bm{a}, \bm {b})$, to obtain Eq.~\eqref{eq:single_k_timeevo}.

\section{Derivation of Eq.~\eqref{eq:h4h4bar}}
\label{sec:h4h4bar}
In this appendix, we derive Eq.~\eqref{eq:h4h4bar}.
We rewrite the operator of interest as
\begin{align}
    H_4-\frac{1}{2}\sum_k a_k \sigma_k \epsilon_k H_2 +\frac{N(N-1)}{4} 
    &=\frac{1}{2} H_2 \cdot \sum_k \left(\frac{1}{2}-a_k\right)\sigma_k \epsilon_k \nonumber \\
    &=\frac{1}{8} \bigg(\sum_k \big(1-a_k+a_k)\sigma_k \epsilon_k \bigg)
    \bigg(\sum_k \big(1-a_k-a_k)\sigma_k \epsilon_k \bigg) \nonumber \\
    &=-\frac{1}{8} \biggl(\sum_k a_k \sigma_k \epsilon_k\biggr)^2 + \frac{1}{8} \biggl(\sum_{k} (1-a_k)\sigma_k \epsilon_k\biggr)^2.
\end{align}
Using this identity together with Eq.~\eqref{eq:sigmaval}, we obtain
\begin{align}
    &\biggl(H_4-\frac{1}{2} \sum_k a_k \sigma_k \epsilon_k H_2 +\frac{N(N-1)}{4}\biggr)\prod_{k} \langle a_k b_k \rangle \nonumber \\
    &=\biggl[-\frac{1}{8} \biggl(\sum_k a_k(-1)^{b_k+1} \epsilon_k \biggr)^2 +\frac{1}{8} \biggl(\sum_{k} (1-a_k)
    (-1)^{b_k+1} \epsilon_k\biggr)^2 \biggr] \prod_k \langle a_k b_k \rangle  \nonumber \\    
    &=\prod_k \langle a_k b_k\rangle \cdot  
    \biggl[-\frac{1}{8} \biggl(\sum_k a_k (-1)^{b_k+1} \epsilon_k \biggr)^2 +\frac{1}{8} \biggl(\sum_{k} (1-a_k)(-1)^{b_k+0} \epsilon_k \biggr)^2 \biggr]  
    \nonumber \\
    &=\prod_k \langle a_k b_k \rangle \cdot \biggl(H_4- \frac{1}{2}\sum_k a_k \sigma_k \epsilon_k H_2 +\frac{N(N-1)}{4} \biggr),
\end{align}
for a given set of $\bm a$ and $\bm b$, which leads to Eq.~\eqref{eq:h4h4bar}.

\section{Orthogonality and completeness of the projection superoperators}
\label{sec:orthogonality}
In this appendix, we prove the orthogonality and completeness of the 
projection superoperators defined in Eq.~\eqref{eq:projection}. 
Throughout this appendix, we assume that the system is away from exceptional points,
so that the matrices $V_k$ are invertible.

We first consider the orthogonality condition,
\begin{equation}
    \Pi_{\bm a;\bm d}\Pi_{\bm a';\bm d'}=\delta_{\bm a, \bm a'}\delta_{\bm d, \bm d'}\Pi_{\bm a;\bm d} \label{eq:orthogonality}.
\end{equation}
For this purpose, we denote 
the coefficient appearing in Eq.~\eqref{eq:projection} by
\begin{align}
    C(\bm a, \bm b,\bm c, \bm d)=
    \prod_{k=1}^{N/2} \biggl[[V_k^{a_{k}}({\bm a,\bm b})]_{c_{k}d_{k}}  [V_{k}^{-a_{k}}({\bm a,\bm b})]_{d_{k }b_{k}} \biggr].
\end{align}
Then, for an arbitrary operator $X$, the action of $\Pi_{\bm a; \bm d}$ is written as
\begin{equation}
    \Pi_{\bm a;\bm d}X 
    =\sum_{ \bm b, \bm c}
    C(\bm a, \bm b,\bm c, \bm d) {\rm Tr} \biggl[ \biggl(\prod_{k=1}^{N/2}\langle a_k b_k \rangle \biggr)^\dagger X \, \biggr]
    \prod_{k=1}^{N/2} \langle a_kc_k\rangle.
\end{equation}

Applying $\Pi_{\bm a; \bm d}\Pi_{\bm p; \bm s}$ to $X$, we obtain
\begin{align}
    \Pi_{\bm a; \bm d}\Pi_{\bm p; \bm s}X=\sum_{\bm b,\bm c,\bm q,\bm r}
    C(\bm a, \bm b, \bm c, \bm d) C(\bm p, \bm q, \bm r, \bm s) 
    {\rm Tr} \biggl[ \biggl(\prod_{k=1}^{N/2}\langle p_k q_k \rangle \biggr)^\dagger X\biggr]
    \cdot {\rm Tr} \biggl[ \biggl(\prod_{k=1}^{N/2}\langle a_k b_k \rangle \biggr)^\dagger \prod_{k=1}^{N/2}\langle p_k r_k \rangle \biggr] \prod_{k=1}^{N/2}\langle a_k c_k \rangle.
\end{align}
Using the orthonormality of the operator basis,
\begin{equation}
    {\rm Tr} \biggl[ \biggl(\prod_{k=1}^{N/2}\langle a_k b_k \rangle \biggr)^\dagger \prod_{k=1}^{N/2}\langle p_k r_k \rangle \biggr]=\delta_{\bm a,\bm p}\delta_{\bm b, \bm r},
\end{equation}
we find
\begin{align}
    \Pi_{\bm a; \bm d}\Pi_{\bm p; \bm s}X
    =\delta_{\bm a,\bm p}\sum_{\bm b,\bm c,\bm q}
    C(\bm a, \bm b, \bm c, \bm d) C(\bm a, \bm q, \bm b, \bm s) 
    {\rm Tr} \biggl[ \biggl(\prod_{k=1}^{N/2}\langle a_k q_k \rangle \biggr)^\dagger X\biggr] \prod_{k=1}^{N/2}\langle a_k c_k \rangle.
    \label{eq:twoprojection}
\end{align}
It remains to evaluate the 
sum over $\bm b$. This 
is carried out as follows:
\begin{align}
    &\sum_{\bm b}C(\bm a, \bm b, \bm c, \bm d) C(\bm a, \bm q, \bm b, \bm s)  \nonumber \\
    &=\sum_{\bm b}\prod_{k=1}^{N/2} \biggl[[V_k^{a_{k}}({\bm a,\bm b})]_{c_{k},d_{k}}  [V_{k}^{-a_{k}}({\bm a,\bm b})]_{d_{k }b_{k}} [V_k^{a_{k}}({\bm a,\bm q})]_{b_{k},s_{k}}  [V_{k}^{-a_{k}}({\bm a,\bm q})]_{s_{k }q_{k}}\biggr] \nonumber \\
    &=\sum_{\bm b}\prod_{k=1}^{N/2} \biggl[[V_k^{a_{k}}({\bm a,\bm q})]_{c_{k},d_{k}}  [V_{k}^{-a_{k}}({\bm a,\bm q})]_{d_{k }b_{k}} [V_k^{a_{k}}({\bm a,\bm q})]_{b_{k},s_{k}}  [V_{k}^{-a_{k}}({\bm a,\bm q})]_{s_{k }q_{k}}\biggr] \nonumber \\
    &=\delta_{\bm d, \bm s} \prod_{k=1}^{N/2}\biggl[[V_k^{a_{k}}({\bm a,\bm q})]_{c_{k},d_{k}} [V_{k}^{-a_{k}}({\bm a,\bm q})]_{d_{k }q_{k}}\biggr] \nonumber \\
    &= \delta_{\bm d, \bm s} C(\bm a, \bm q, \bm c, \bm d).
    \label{eq:Csummation}
\end{align}
In deriving Eq.~\eqref{eq:Csummation}, we have replaced $V_k^{a_k}({\bm a,\bm b})$ and $V_k^{-a_k}({\bm a,\bm b})$ by $V_k^{a_k}({\bm a,\bm q})$ and $V_k^{-a_k}({\bm a,\bm q})$, respectively. This replacement is justified as follows.

Let $\mathcal{I}_{N/2}=\{1,\dots,N/2\}$. 
For an arbitrary $k^*\in \mathcal{I}_{N/2}$, 
we consider the substitution $b_{k_*}\to q_{k_*}$ in $V_k^{a_k}({\bm a,\bm b})$.
We also define $\bm b'=(b_1,\dots,q_{k_*}, \dots,b_{N/2})$ and
separate the argument into three cases:

\medskip
\begin{enumerate}[label=(\roman*)]
\item For $k=k_*$, the matrix $V_{k}^{a_k}(\bm a,\bm b)$ is independent of $b_{k_*}$. 

\item For $k\neq k_*$ and $a_{k_*}=1$, the matrix $V_{k}^{a_k}(\bm a,\bm b)$ is also independent.

\item For $k\neq k_*$ and $a_{k_*}=0$, the relevant factor becomes 
\begin{equation}
[V_{k_*}^0(\bm a,\bm b)]_{c_{k_*} d_{k_*}}[V_{k_*}^{0}(\bm a,\bm b)]_{d_{k_*} b_{k_*}}[V_{k_*}^{0}(\bm a,\bm q)]_{b_{k_*} s_{k_*}}[V_{k_*}^{0}(\bm a,\bm q)]_{s_{k_*} q_{k_*}}=\delta_{c_{k_*} d_{k_*}}\delta_{d_{k_*} b_{k_*}} \delta_{b_{k_*} s_{k_*}} \delta_{s_{k_*} q_{k_*}}.
\end{equation}
This factor enforces $b_{k_*}=c_{k_*}=d_{k_*}=s_{k_*}=q_{k_*}$. 
\end{enumerate}
Therefore, the replacement $V_{k}^{a_k}(\bm a,\bm b) \to V_{k}^{a_k}(\bm a,\bm b')\, $ for all $k\in \mathcal{I}_{N/2}$ is justified. 
Since $k_*$ is arbitrary, we may repeat this argument for all components 
and replace $\bm b$ by $\bm q$.
This proves Eq.~\eqref{eq:Csummation}. Combining Eqs.~\eqref{eq:twoprojection} and \eqref{eq:Csummation}, we obtain Eq.~\eqref{eq:orthogonality}.

Next, we prove the completeness condition. Summing $C({\bm a},{\bm b},{\bm c},{\bm d})$ over $\bm d\in \{0,1\}^{N/2}$ gives $\prod_{k=1}^{N/2}\delta_{c_k,b_k}$.
Using this relation together with the completeness of the operator CONS, we obtain
\begin{equation}
    \sum_{\bm a, \bm d} \Pi_{{\bm a; \bm d}}X=\sum_{\bm a,\bm b, \bm c}\prod_{k=1}^{N/2}\delta_{c_k,b_k} {\rm Tr} \biggl[ \biggl(\prod_{k=1}^{N/2}\langle a_k b_k \rangle \biggr)^\dagger X \, \biggr] \prod_{k=1}^{N/2} \langle a_kc_k\rangle=X. 
    \label{eq:completeness}
\end{equation}
Thus, the projection superoperators are complete.

\end{widetext}

\section{Derivation of Eqs.~\eqref{eq:analytics} and \eqref{eq:phi_asymptotics}\label{sec:correlation_analytics}}
In this appendix, we derive Eqs.~\eqref{eq:analytics} and
\eqref{eq:phi_asymptotics} 
in the large-$N$ limit at fixed
$\xi=N\Gamma$ and $t$. We first rewrite $C^{i,i}_{\infty}(t)$ as
\begin{align}
C^{i,i}_{\infty}(t)=\frac{2e^{-\xi t}}{N}\sum_{k=1}^{N/2}\prod_{n\neq k} Q_{n,k}(t),
\end{align}
where
\begin{align}
Q_{n,k}(t)=\cosh(G_{n,k}t)-\frac{2\Gamma}{G_{n,k}}\sinh(G_{n,k}t).
\end{align}
The dominant contribution in the large-$N$ limit comes from the modes close to
$k=N/2$. Indeed, consider modes that remain a macroscopic distance
from $k=N/2$, namely,
\begin{align}
\frac{N}{2}-k\geq\delta N
\end{align}
for some $N$-independent $0<\delta<1/2$. For these modes,
\begin{align}
\epsilon_k&=2\cot\left(\frac{(2k-1)\pi}{2N}\right)
\nonumber\\
&\geq 2\tan\left(\delta\pi+\frac{\pi}{2N}\right)>2\tan(\delta\pi)>0.
\end{align}
Thus, $\epsilon_k$ is bounded below by an $N$-independent positive
constant. For the modes satisfying $n/N\to y\in(0,1/2)$, one has
\begin{equation}
\epsilon_n\to 2\cot(\pi y).
\end{equation}
Since $\Gamma=\xi/N\to0$, they satisfy
\begin{align}
G_{n,k}=\imi\frac{\epsilon_n\epsilon_k}{2}+o(1),
\quad \frac{2\Gamma}{G_{n,k}} = o(1).
\end{align}
Consequently, for fixed $t>0$,
\begin{align}
Q_{n,k}(t) = \cos\left(\frac{\epsilon_n\epsilon_k t}{2}\right)+o(1).
\end{align}
For any such $k$ and fixed $t$, we can take an interval $I_{x,t}$ of nonzero length and a constant $0<\eta<1$ such that, for all sufficiently large $N$,
\begin{equation}
|Q_{n,k}(t)|<\eta
\end{equation}
whenever $n/N\in I_{x,t}$. The number of such modes is
\begin{equation}
|I_{x,t}|N+o(N).
\end{equation}
Consequently,
\begin{align}
\left| \prod_{n\neq k}Q_{n,k}(t) \right| = O\left(e^{-cN}\right),
\end{align}
where $c=c(x,t)>0$. Hence, such modes do not contribute to the leading
large-$N$ asymptotics. We therefore set
\begin{align}
k=\frac{N}{2}-m,\quad m=0,1,2,\ldots.
\end{align}
Then
\begin{align}
\theta_k=\pi-\frac{(2m+1)\pi}{N},
\end{align}
and hence
\begin{align}
\epsilon_k
&=2\cot\frac{\theta_k}{2}
=2\tan\frac{(2m+1)\pi}{2N}\nonumber\\
&=\frac{(2m+1)\pi}{N}+O(N^{-3}).
\end{align}

On the other hand, for fixed $n=O(1)$,
\begin{align}
\epsilon_n
=2\cot\frac{(2n-1)\pi}{2N}
=\frac{4N}{(2n-1)\pi}+O(N^{-1}).
\end{align}
Therefore, 
\begin{align}
\frac{\epsilon_n\epsilon_k}{2}
=\frac{2(2m+1)}{2n-1}+O(N^{-2}).
\end{align}
Since $\Gamma=\xi/N$, we obtain
\begin{align}
G_{n,k}=\sqrt{4\Gamma^2-\frac{(\epsilon_n\epsilon_k)^2}{4}}
=\imi\frac{\epsilon_n\epsilon_k}{2}+O(N^{-2}).
\end{align}
Since 
\begin{equation}
\frac{2\Gamma}{G_{n,k}}=O(N^{-1}),
\end{equation}
the $\sinh$ term in $Q_{n,k}(t)$ can be ignored for fixed $n=O(1)$. Therefore, 
\begin{align}
Q_{n,k}(t)
=\cos\left(
\frac{2(2m+1)t}{2n-1}
\right)+O(N^{-1}).
\end{align}

However, the $\sinh$ term cannot simply be discarded in the full product. It is
small for each fixed $n$, but 
taking the product over $O(N)$ modes yields an additional exponential factor. For bulk modes with $n=O(N)$, $\epsilon_n=O(1)$. Hence
\begin{align}
\epsilon_n\epsilon_k=O(N^{-1}),\quad \Gamma=\frac{\xi}{N}=O(N^{-1}),
\end{align}
and therefore
\begin{align}
G_{n,k} = \sqrt{ 4\Gamma^2-\frac{(\epsilon_n\epsilon_k)^2}{4} } = O(N^{-1}).
\end{align}
For fixed $t$, this gives
\begin{align}
Q_{n,k}(t) &= \cosh(G_{n,k}t) -\frac{2\Gamma}{G_{n,k}}\sinh(G_{n,k}t) \nonumber\\
&=1-\frac{2\xi t}{N}+O(N^{-2}).
\end{align}
Hence
\begin{align}
\prod_{n\neq k} Q_{n,k}(t) = e^{-\xi t} \prod_{n=1}^{\infty} \cos\left( \frac{2(2m+1)t}{2n-1} \right)+o(1).
\end{align}
Defining
\begin{align}
\Phi(u) = \prod_{r=0}^{\infty} \cos\left( \frac{2u}{2r+1} \right)=\cos(2u)\prod_{r=1}^{\infty} \cos\left( \frac{2u}{2r+1} \right),
\end{align}
we arrive at
\begin{align}
C^{i,i}_{\infty}(t) = \frac{2e^{-2\xi t}}{N} \sum_{m=0}^{\infty} \Phi\bigl((2m+1)t\bigr)+o(N^{-1}).
\end{align}
This gives Eq.~\eqref{eq:analytics}.

We next estimate the large-$u$ envelope of $\Phi(u)$. By decomposing
\begin{align}
   \Phi(u) &= \cos(2u) \Psi(u),\\
   \Psi(u)&=\prod_{r=1}^{\infty} \cos\left( \frac{2u}{2r+1} \right), 
\end{align}
and taking the logarithm,
\begin{align}
\log|\Psi(u)| = \sum_{r=1}^{\infty} \log\left| \cos\left(\frac{2u}{2r+1}\right) \right|.
\end{align}
For large $u$, we approximate the sum by a continuum integral. Introducing
$y=\frac{2r+1}{2u}$, whose spacing is $1/u$, we obtain
\begin{align}
\log|\Psi(u)| \simeq u \int_0^\infty \log\left| \cos\frac{1}{y} \right| dy.
\end{align}
Changing variables to $z=1/y$ gives
\begin{align}
\log|\Psi(u)| \simeq u \int_0^\infty \frac{dz}{z^2} \log|\cos z|.
\end{align}

A natural way to evaluate the integral is to use the standard Fourier
expansion
\begin{align}
\log|\cos z|=-\log 2+\sum_{s=1}^{\infty}(-1)^{s+1}\frac{\cos(2sz)}{s}.
\end{align}
Since
\begin{align}
\log 2 = \sum_{s=1}^{\infty} (-1)^{s+1}\frac{1}{s},
\end{align}
this expansion can equivalently be written as
\begin{align}
\log|\cos z| = \sum_{s=1}^{\infty} (-1)^{s+1}\frac{1}{s} \left[\cos(2sz)-1\right].
\end{align}
However, this conditionally convergent Fourier series cannot be integrated term by term in the present integral. A formal interchange of the sum and the integral would give
\begin{align}
&\int_0^\infty \frac{dz}{z^2}\log|\cos z|\nonumber\\
&\stackrel{\mathrm{formal}}{=}\sum_{s=1}^{\infty}
(-1)^{s+1}\frac{1}{s}
\int_0^\infty dz\,
\frac{\cos(2sz)-1}{z^2}
\nonumber\\
&=-\pi\sum_{s=1}^{\infty}(-1)^{s+1},
\end{align}
where we used
\begin{align}
\int_0^\infty dz\,
\frac{\cos(2sz)-1}{z^2}
=-\pi s.
\end{align}
The resulting series \(1-1+1-\cdots\) does not converge in the ordinary
sense. Thus, although the original integral is convergent, the above
term-by-term integration is not justified.

To make this manipulation well defined, we use Abel summation
\cite{Stein_Shakarchi_2003}. For \(0<\rho<1\), we introduce the
Abel-regularized Fourier series
\begin{align}
F_\rho(z)= \sum_{s=1}^{\infty} (-1)^{s+1} \frac{\rho^s}{s} \left[\cos(2sz)-1\right].
\end{align}
Then
\begin{align}
F_\rho(z) = \frac{1}{2} \log\left(1+2\rho\cos 2z+\rho^2\right) -\log(1+\rho),
\end{align}
and $F_\rho(z)\to \log|\cos z|$ as $\rho\to1^-$. For $\rho<1$, the sum and
the integral can be interchanged, giving
\begin{align}
\int_0^\infty \frac{dz}{z^2}F_\rho(z)
&=\sum_{s=1}^{\infty} (-1)^{s+1} \frac{\rho^s}{s} \int_0^\infty dz
\frac{\cos(2sz)-1}{z^2}\nonumber\\
&=-\pi \sum_{s=1}^{\infty} (-1)^{s+1}\rho^s
=-\pi\frac{\rho}{1+\rho}.
\end{align}
Taking $\rho\to1^-$, we obtain
\begin{align}
\int_0^\infty \frac{dz}{z^2} \log|\cos z| =-\frac{\pi}{2}.
\end{align}

Therefore,
\begin{align}
\log|\Psi(u)| \simeq -\frac{\pi}{2}u,
\end{align}
or equivalently,
\begin{align}
|\Phi(u)|\simeq \exp\left(-\frac{\pi}{2}u\right)\cos(2u).
\end{align}
This gives Eq.~\eqref{eq:phi_asymptotics}.

\section{Derivation of Eqs.~\eqref{eq:noncommute1} and \eqref{eq:noncommute2}\label{sec:singularity}}
\begin{figure}
 \centering
  \includegraphics[width=\linewidth]{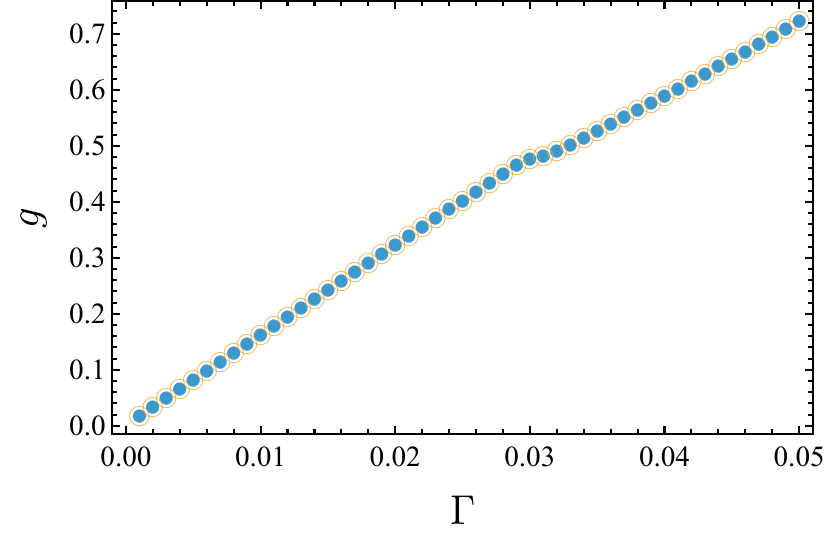}
  \caption{The asymptotic decay rate $g$ as a function of $\Gamma$ ($N=16$). Blue points are calculated with Eq.~\eqref{eq:asymptotic_decay} and orange circles are calculated with Eq.~\eqref{eq:asymptotic_decay_2}.}
  \label{fig:gap_16}
\end{figure}

In this appendix, we derive Eqs.~\eqref{eq:noncommute1} and \eqref{eq:noncommute2} in the main text. First, it is numerically confirmed that $\operatorname{Re}\lambda_{l,\bm{b}}$ takes the maximum when $l=N/2$ and $b_k=0$ for $k\neq l$ (See Fig.~\ref{fig:gap_16}). 
From this, we assume for any $\Gamma$ and $N$ that
\begin{equation}
    g=-\operatorname{Re}\left[-N\Gamma+2\sum_{k=1}^{N/2-1} \sqrt{\Gamma^2-[f(k)f(N/2)]^2}\right],
    \label{eq:asymptotic_decay_2}
\end{equation}
where $f(k)=\cot\left[\frac{(2k-1)\pi}{2N}\right]$.
First, for any $N$, we can see that 
\begin{equation}
    \lim_{\Gamma\to 0}g=-\operatorname{Re}\left[\sum_{k=1}^{N/2-1} \sqrt{-[f(k)f(N/2)]^2}\right]=0,
\end{equation}
and therefore 
\begin{equation}
    \lim_{N\to \infty}\lim_{\Gamma\to 0}g=0.
\end{equation}
On the other hand, $\lim_{\Gamma\to 0}\lim_{N\to \infty}g\neq0$. To see this, we first fix $\Gamma>0$ and take the limit $N\to \infty$ first:
\begin{equation}
    \lim_{N\to \infty}g=2\Gamma-\lim_{N\to \infty}\sum_{k=1}^{N/2} 2X(k),
    \label{eq:sum_0}
\end{equation}
where
\begin{equation}
    X(k)=\operatorname{Re}\left[\sqrt{\Gamma^2-[f(k)f(N/2)]^2}-\Gamma\right].
\end{equation}
Here, we used that $\lim_{N\to \infty} X(N/2) = 0$. Note that, $f(k)$ is a monotonically decreasing function with respect to $k$ and for fixed $k$,
\begin{align}
    \lim_{N\to\infty}f(k)f(N/2)
    &=\lim_{N\to\infty}\cot\left[\frac{(2k-1)\pi}{2N}\right]\tan\left[\frac{\pi}{2N}\right]\nonumber\\
    &=\frac{1}{2k-1}.
    \label{eq:limit_evaluation}
\end{align}

To evaluate $X(k)$, we consider cases based on the sign inside the square root. If $f(k)f(N/2)>\Gamma$,
we can see that
\begin{equation}
    X(k)=-\Gamma.
\end{equation}
From Eq.~\eqref{eq:limit_evaluation}, the largest $k_0$ such that $f(k_0)f(N/2)>\Gamma$ should satisfy
\begin{equation}
    \frac{1}{2k_0+1}<\Gamma<\frac{1}{2k_0-1},
\end{equation}
which can be rewritten as 
\begin{equation}
    \frac{1-\Gamma}{2\Gamma}<k_0<\frac{1+\Gamma}{2\Gamma}.
\end{equation}
Thus, 
\begin{equation}
   \frac{-1-\Gamma}{2} <\lim_{N\to \infty}\sum_{k=1}^{k_0} X(k)<\frac{-1+\Gamma}{2}
\end{equation}
and therefore
\begin{equation}
   \lim_{\Gamma\to 0}\lim_{N\to \infty}\sum_{k=1}^{k_0} X(k)=-\frac{1}{2}.
    \label{eq:sum_1}
\end{equation}
If $f(k)f(N/2)\leq \Gamma$, we have
\begin{equation}
    X(k)=\sqrt{\Gamma^2-[f(k)f(N/2)]^2}-\Gamma.
\end{equation}
Therefore,
\begin{align}
   &\lim_{N\to \infty}\sum_{k=k_0+1}^{N/2} X(k)\nonumber\\
   &=\lim_{N\to \infty}\! \frac{1}{N}\!\!\sum_{k=k_0+1}^{N/2} \!\left[\sqrt{(N\Gamma)^2-\frac{\pi^2}{4}\cot^2\left[\frac{(2k-1)\pi}{2N}\right]}-N\Gamma\right]\nonumber\\
   &=\lim_{N\to \infty} \int_{\frac{{k_0}+1}{N}}^\frac{1}{2}\left[\sqrt{(N\Gamma)^2-\frac{\pi^2}{4}\cot^2\left(\pi x\right)}-N\Gamma\right]dx\nonumber\\
   &= \lim_{N\to \infty} \int_{\Gamma(k_0+1)}^{\frac{N\Gamma}{2}}\left[\sqrt{1-\frac{1}{4y^2}}-1\right]dy,
\end{align}
where we set $x=\frac{k}{N}$ and $y=N\Gamma x$. To evaluate this integral, we consider the following indefinite integral, 
\begin{align}
    \int \sqrt{1-\frac{1}{4y^2}} dy
    &=\int \tan^2\theta d\theta \nonumber\\
    &=\int (\sec^2\theta-1) d\theta \nonumber\\
   &=\frac{1}{2}(\tan \theta-\theta)+C\nonumber\\
    &= \sqrt{y^2-\frac{1}{4}}-\frac{1}{2}\arcsec (2y)+C,
\end{align}
where we set $y=\frac{1}{2}\sec\theta$ and $C$ is a constant of integration.  Additionally, in the limit as $\Gamma\to 0$, the lower limit of integration evaluates to
\begin{equation}
    \lim_{\Gamma\to 0}\Gamma(k_0+1)=\frac{1}{2}.
\end{equation}
Therefore, by applying these results to compute the definite integral and taking the limits, we finally obtain:
\begin{align}
   &\lim_{\Gamma\to 0}\lim_{N\to \infty}\sum_{k=k_0+1}^{N/2} X(k)\nonumber\\
   &=\lim_{\Gamma\to 0}\lim_{N\to \infty}\left[\sqrt{y^2-\frac{1}{4}}-\frac{1}{2}\arcsec(2y)-y\right]_{y=\frac{1}{2}}^{\frac{N\Gamma}{2}}\nonumber\\
   &=-\frac{\pi}{4}+\frac{1}{2}.
   \label{eq:sum_2}
\end{align}

From Eqs.~\eqref{eq:sum_0}, \eqref{eq:sum_1} and \eqref{eq:sum_2}, we have
\begin{equation}
    \lim_{\Gamma\to 0}\lim_{N\to \infty}g=\frac{\pi}{2}.
\label{eq:noncommute2_appendix}
\end{equation}

\begin{figure}
 \centering
  \includegraphics[width=\linewidth]{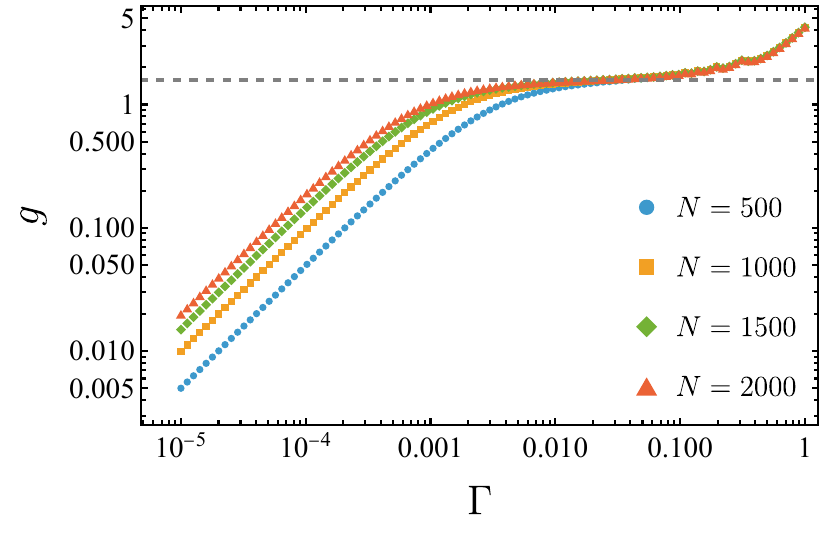}
  \caption{The asymptotic decay rate $g$ as a function of $\Gamma$ ($N=500,1000,1500,2000$) calculated with Eq.~\eqref{eq:asymptotic_decay_2}. The gray dashed line indicates the line $g=\frac{\pi}{2}$. }
  \label{fig:gap_limit}
\end{figure}

This result is also numerically confirmed. In Fig.~\ref{fig:gap_limit}, $g$ is plotted as a function of $\Gamma$. For fixed $N$, we can see that $\lim_{\Gamma\to 0} g$ converges to zero. On the other hand, for fixed $\Gamma$, $\lim_{N\to\infty} g$ converges to a value which is close to $\frac{\pi}{2}$.

\bibliography{reference}
\end{document}